\documentclass[epsf,prb,showpacs,nofootinbib,10pt]{revtex4-2}
\pdfoutput=1

\usepackage{amsmath,amssymb,bm,hhline} 
\usepackage{makecell}
\usepackage{graphicx}
\usepackage[caption=false]{subfig}
\usepackage{color} 
\usepackage[papersize={8.5in,11in}]{geometry}
\usepackage[colorlinks=true]{hyperref}  
\hypersetup{
    bookmarks=true,         % show bookmarks bar?
    unicode=false,          % non-Latin characters 
    pdftoolbar=true,        % show Acrobat
    pdfmenubar=true,        % show Acrobat 
    pdffitwindow=false,     % window fit to page when opened
    pdfstartview={FitH},    % fits the width of the page to the window
    pdftitle={My title},    % title
    pdfauthor={Author},     % author
    pdfsubject={Subject},   % subject of the document
    pdfcreator={Creator},   % creator of the document
    pdfproducer={Producer}, % producer of the document
    pdfkeywords={keyword1} {key2} {key3}, % list of keywords
    pdfnewwindow=true,      % links in new window
    colorlinks=true,       % false: boxed links; true: colored links
    linkcolor=magenta, %red,          % color of internal links (change box color with linkbordercolor)
    citecolor=blue,        % color of links to bibliography
    filecolor=magenta,      % color of file links
    urlcolor=blue           % color of external links
} 

\usepackage{amsfonts}
\usepackage{upgreek}
\usepackage{slashed}
\usepackage{latexsym}
\usepackage{ulem}

\usepackage{wasysym}			
\usepackage{amsthm}
\usepackage{xcolor}

\usepackage{cancel} % ructrike through
\usepackage{eucal} % The package changes the font that is selected by \mathcal
\usepackage{feynmp}    %Package for feynman diagrams.
\usepackage{mathrsfs}%\mathscr
\usepackage{soul}
\usepackage{wrapfig} % to include figure with text wrapping around it

\newcommand{\nn}{\nonumber \\}
\newcommand{\be}{\begin{equation}}
\newcommand{\ee}{\end{equation}}
\newcommand{\diag}{\mathrm{diag}}
\newcommand\dvec{{\boldsymbol{d}}}
\newcommand{\Tr}{{\rm Tr}\,}
\newcommand{\bp}{{\vec{p}}}
\newcommand{\bq}{{\vec{q}}}
\newcommand{\addRN}[1]{\textcolor{red}{#1}}

\renewcommand{\epsilon}{\varepsilon}
\renewcommand{\vec}[1]{{\bf #1}}

\begin{document}
\title{Interplay of Coulomb interactions and disorder in three dimensional quadratic band crossings without time-reversal symmetry and with unequal masses for conduction and valence bands}
\date{\today}

\author{Ipsita Mandal}
\affiliation{Department of Physics, Indian Institute of Technology, Kharagpur, 721302 India}

\author{Rahul M. Nandkishore}
\affiliation{Department of Physics and Center for Theory of Quantum Matter, University of Colorado at Boulder, Boulder, CO 80309, USA}

\begin{abstract}
Coulomb interactions famously drive three dimensional quadratic band crossing semimetals into a non-Fermi liquid phase of matter. In a previous work, {\it Phys. Rev. B 95, 205106 (2017)}, the effect of disorder on this non-Fermi liquid phase was investigated, assuming that the bandstructure was isotropic, assuming that the conduction and valence bands had the same band mass, and assuming that the disorder preserved exact time-reversal symmetry and statistical isotropy. It was shown that the non-Fermi liquid fixed point is unstable to disorder, and that a runaway flow to strong disorder occurs. In this work, we extend that analysis by relaxing the assumption of time-reversal symmetry and allowing the electron and hole masses to differ (but continuing to assume isotropy of the low energy bandstructure). We first incorporate time-reversal symmetry breaking disorder, and demonstrate that there do not appear any new fixed points. Moreover, while the system continues to flow to strong disorder, time-reversal-symmetry-breaking disorder grows asymptotically more slowly than time-reversal-symmetry-preserving disorder, which we therefore expect should dominate the strong-coupling phase. We then allow for unequal electron and hole masses. We show that whereas asymmetry in the two masses is {\it irrelevant} in the clean system, it is {\it relevant} in the presence of disorder, such that the `effective masses' of the conduction and valence bands should become sharply distinct in the low-energy limit. We calculate the RG flow equations for the disordered interacting system with unequal band masses, and demonstrate that the problem exhibits a runaway flow to strong disorder. Along the runaway flow, time-reversal-symmetry-preserving disorder grows asymptotically more rapidly than both time-reversal-symmetry-breaking disorder and the Coulomb interaction. 
\end{abstract}

\maketitle
\tableofcontents

%==========================================================================

\section{Introduction}
In 1971 Abrikosov studied isotropic three dimensional systems with {\it quadratic} band crossings using a renormalization group calculation in $4-\epsilon$ dimensions \cite{Abrikosov}, and argued  that Coulomb interactions could stabilize a {\it non-Fermi liquid}  phase.
 Interest in this problem has recently been revived ~\cite{MoonXuKimBalents, Herbut, Herbut2, Herbut3, LABIrridate} because of its relevance for pyrochlore iridates. However, theoretical explorations have largely been confined to the {\it clean} (disorder-free) problem, whereas realistic materials are always disordered to some degree. The interplay of disorder with Coulomb interactions in three dimensional quadratic band crossings is a particularly rich problem, since both disorder and Coulomb interactions are {\it relevant} with the {\it same} scaling dimension, and thus  should be treated on an equal footing. 

In two recent works ~\cite{LaiRoyGoswami, rahul-sid} the interplay of disorder and Coulomb interactions was investigated, assuming (a) exact time-reversal symmetry (b) equal band masses for the electron and hole bands and (c) isotropy. It was shown that disorder is a relevant perturbation to Abrikosov's non-Fermi liquid fixed point, and that the disordered problem undergoes a runaway flow to strong disorder, the implications of which were discussed at length in Ref.~\cite{rahul-sid}.  In this work, we extend the analysis (working with the renormalization group scheme of  Ref.~\cite{rahul-sid}) to incorporate time-reversal symmetry breaking disorder (which may arise physically from e.g. magnetic impurities), and allowing also for unequal band masses of the conduction and valence bands. We have used the powerful technique of dimensional regularization, which has been succesfully used in many problems involving non-fermi liquids \cite{nfl0,nfl1,*nfl2,*nflsc}. %We show that the problem continues to exhibit a runaway flow to strong disorder, and no new fixed points appear. We emphasize that PH symmetry-breaking, while an {\it irrelevant} perturbation in the clean problem, is a {\it relevant} perturbation to the disordered problem. This provides a clean experimental diagnostic for the effect of disorder in quadratic band crossing materials. 

This paper is structured as follows. In Sec.~\ref{model} we introduce the basic model and renormalization group scheme. In Sec.~\ref{sec:ph-sym} we calculate the interplay of interactions and disorder, including time-reversal symmetry breaking `tensor' disorder which was ignored in previous analyses. We show that no new fixed points appear, and the problem continues to flow to strong disorder. We further show that time-reversal symmetry breaking disorder grows asymptotically more slowly than time-reversal symmetry preserving disorder as the problem flows to strong disorder. In Sec.~\ref{sec:ph-asym} we further relax the assumption of equal electron and hole masses. We find that asymmetry in the masses is a {\it relevant} perturbation to the disordered system, and comes to dominate the low energy physics. We find this result surprising, because such asymmetry is irrelevant in the clean system \cite{Abrikosov}. (For another setting where such asymmetry affects disorder physics, see Ref.~\cite{Weichman}).  The interplay of disorder and interactions must thus be re-analyzed in the presence of unequal masses - a task we perform. Our analysis of the $\beta$ functions, and discussion of the results, are presented in Sec.~\ref{sec:analysis}. 
The appendices contain technical results employed in the derivations, but which are inessential to the flow of the argument.

\section{\label{model}Model}

We consider a model for three-dimensional quadratic band crossings, where the low energy bands form %lie in
 a four-dimensional representation of the lattice symmetry group \cite{MoonXuKimBalents}. Then the $\vec{k} \cdot \vec{p}$ Hamiltonian for the non-interacting system, in the absence of disorder, takes the form \cite{Herbut}:
 \begin{equation}
 \mathcal{H}_0 = \sum_{a=1}^N d_a(\vec{k}) \,  \Gamma_a   + \frac{k^2}{2\,m'} \,,
\quad d_a  (\vec{k})   = \frac{\tilde d_a(\vec{k})   }{2\, m}\,,
\label{bare}
 \end{equation}
%%%%%%%%%%%%%%%%
where the $\Gamma_a$'s are the {rank four irreducible representations of the Clifford algebra relation $\{\Gamma_a,\Gamma_b\} = 2\, \delta_{ab}$ in the Euclidean space. We have used the usual notation $\{A,B\} = AB+BA$ for denoting the anticommutator. There are $N=5$ such matrices, which are related to the familiar gamma matrices from the Dirac equation (plus the matrix conventionally denoted as $\gamma_5$), but with the Euclidean metric $\{\Gamma_a,\Gamma_b\} = 2\, \delta_{ab}$ instead of the Minkowski metric $\{\Gamma_a,\Gamma_b\} = 2\, (-1,+1,+1,+1)$. Using various Clifford algebra relations, as shown in Appendix~\ref{app1}. In $d=3$, the space of $4\times 4$ Hermitian matrices is spanned by the identity matrix, the five $4\times 4$ Gamma matrices $\Gamma_a$ and the ten distinct matrices $\Gamma_{ab} = \frac{1}{2i}\, [\Gamma_a, \Gamma_b]$.
Furthermore, the  $\tilde d_a(\vec k)$'s are the $l=2$ spherical harmonics, which have the following structure:
\begin{align}
\label{ddef}
\tilde d_1(\vec{k}) &= \sqrt{3}\, k_y \,k_z\,,
\quad \tilde d_2(\vec{k}) =  \sqrt{3}\, k_x\, k_z\, ,\quad
 \tilde d_3(\vec{k}) =  \sqrt{3} \,k_x\, k_y\, ,\quad
 \tilde d_4(\vec{k}) =\frac{\sqrt{3}  \,  (k_x^2 - k_y^2) }{2}\,, \quad
 \tilde d_5(\vec{k}) = \frac{2\, k_z^2 - k_x^2 - k_y^2}{2} \,.
\end{align}
The isotropic $\frac{k^2}{2\,m'}$ term with no spinor structure makes the band mass of the conduction and valence bands unequal. This is an irrelevant perturbation in the clean system \cite{Abrikosov}, and was ignored in previous analyses \cite{LaiRoyGoswami, rahul-sid}. We consider a setting with $N_f$ independent flavours of fermions.

\subsection{Action}

 The full action was derived in Ref.~\cite{rahul-sid}, and takes the form:  
%%%%%%%%%%%
\begin{align}
S =& \sum  \limits_{i,\xi}  \int d\tau\, d^d x\, { \psi_i^\xi}^{\dag}(x,\tau)\, [\, \partial_{\tau} + \mathcal{H}_0(x)\, ]\,  {\psi_i^\xi}(x,\tau) 
\nn  & + \sum  \limits_{i,\xi , \xi'} 
 \frac{e^2}{2 \,c } \int d\tau \,d^d q\, d^d p\, d^d p'\, 
V(q) \,{ \psi_i^\xi}^{\dag}(\vec{p},\tau)\, { \psi_i^{\xi'}}^{\dag}(\vec{p'},\tau) \,\psi_{i}^{\xi'} (\vec{p'-q},\tau) \,\psi_{i}^{\xi} (\vec{p+q},\tau)
 \nonumber \\ 
& -  W_0 \sum \limits_{i,j,\xi}   \int d\tau\, d\tau' \,d^d x \,({ \psi_i^\xi}^{\dag} \, {\psi_i^\xi})_{\tau} \,({ \psi_j^\xi}^{\dag} \, \psi_j^\xi  )_{\tau'} 
- W_1 \sum  \limits_{i,j, \xi } \sum \limits_{a}  \int d\tau \,d\tau' \,d^d x\, ({ \psi_i^\xi}^{\dag}\, \Gamma_a \, {\psi_i^\xi})_{\tau} \,({ \psi_j^\xi}^{\dag} \,\Gamma_a\, \psi_j^\xi )_{\tau'}\nn
& - W_2 \sum  \limits_{i,j,\xi}  
 \sum \limits_{a<b}\int d\tau \,d\tau' \,d^d x\, ({ \psi_i^\xi}^{\dag} \, \Gamma_{ab} \, {\psi_i^\xi})_{\tau} \,({ \psi_j^\xi}^{\dag}\, \Gamma_{ab}\, \psi_j^\xi)_{\tau'}\,. \label{fullactionMoon}
\end{align}
%%%%%%%%%%%%%%%%%%%%
%{Coulomb line should couple different flavors}
Here the Coulomb interaction $V(q) = \frac{1}{q^2}$ has been written in momentum space, and $\xi =(1,2,\ldots,N_f)$ (same with $\xi'$) denotes the flavour index. Note that disorder has been taken to be diagonal in the flavor space. 
The disorder terms, parametrized by the constants $W_\alpha$ ($\alpha=0,1,2$), represent short-range-correlated disorder with and without spinor structure. The disorder  is treated in the replica formalism with replica indices $i, j$, with $n$ replicas.  The limit $n\rightarrow 0$ has to be taken at the end of the computation. The sums over $a$ range over all the five independent $4\times 4$ non-identity Hermitian matrices $\Gamma_a$ in the spinor space, while the sums over $(a,b)$ range over all the ten independent matrices $\Gamma_{ab}$. Here $W_0$ represents a random chemical potential, and $W_2$ represents magnetic disorder, which breaks time reversal symmetry. Meanwhile, $W_1$ represents `vector' disorder (e.g. strain) which preserves all the symmetries of the problem (apart from translation symmetry). For a further discussion of the different disorder types, see Ref.\cite{rahul-sid}. Furthermore, the short-ranged interactions have been neglected as they are less relevant in an RG sense, compared to either the long-ranged (Coulomb) interactions or short-range-correlated disorder.

The Hamiltonian in Eq.~(\ref{bare}), in the absence of the $\frac{k^2}{2\,m'}$ term, has been considered in the presence of  ``scalar" (disorder vertex contains no gamma matrix) and ``vector" (disorder vertex contains one gamma matrix) disorder terms, in the absence and presence of the Coulomb interaction, in Ref.~\cite{rahul-sid}. In Sec.~\ref{sec:ph-sym} of the present work we generalize the analysis to include time reversal symmetry breaking `tensor' disorder ($W_2 \neq 0$). Meanwhile, the $\frac{k^2}{2\,m'}$ term was dropped in Ref.~\cite{rahul-sid} as it is irrelevant in the $\vec{k} \cdot \vec{p}$ Hamiltonian in the presence of Coulomb interactions \cite{Abrikosov, MoonXuKimBalents} for the clean system. In Sec.~\ref{sec:ph-asym} we re-introduce this term, and analyze the interplay of all kinds of disorder and Coulomb interactions allowing for unequal band masses.

%%%%%
\subsection{Scaling dimensions and RG scheme}

%Let us denote the fermion field in the energy-momentum space by $\tilde \psi^\xi$.
The canonical scaling dimensions are given by: $[x^{-1}] = 1$ and $[\tau^{-1}] = z$ (dynamical critical exponent). From the invariance of the bare action, we find that $[\psi^\xi] = d/2$ and $[m]=[m']= 2-z$ in $d$ spatial dimensions, where $z=2$ at tree level. For the Coulomb and disorder terms, we get:
\begin{equation}
[e^2] = z+2-d \,,
\qquad [W_{\alpha}] = 2\,z - d - 2\, \eta_{\alpha}  \,,
\label{dimensions}
\end{equation}
where we have allowed for the anomalous exponents $[  { \psi^\xi}^{\dag}\,  \psi^\xi] = d + \eta_{0}$, $[  {\psi^\xi}^{\dag}\, \Gamma_a\, \psi^\xi] = d + \eta_{1}$ and $[{\psi^\xi}^{\dag}\, \Gamma_{ab}\,  \psi^\xi ] = d + \eta_{2}$. The anomalous exponents are zero at the Gaussian fixed point, and hence all the $W_{\alpha}$'s have the same tree-level scaling. Since both the Coulomb interactions and disorder are relevant at tree level with the {\it same} exponent, at least at the Gaussian fixed point about which the perturbation is being carried out, they must be treated on an equal footing. 

Our RG scheme involves a continuation to $d=4-\epsilon$ spatial dimensions. In $d=4$, the Coulomb interaction and disorder are marginal at tree level, and controlled calculations may be carried out at small $\epsilon$. (Of course, a description of the physical situation in $d=3$ requires a continuation to $\epsilon = 1$, which could be problematic \cite{Herbut}). However, the extension to four dimensions employed in the classic analysis of Abrikosov \cite{Abrikosov} is unsuitable for two reasons. Firstly, it greatly expands the number of $\Gamma$ matrices in the problem, leading to the introduction of disorder types that have no analog in the physical problem in $d=3$. Furthermore, Abrikosov's dimensional continuation changes the representation of time reversal from $T^2 = -1$ (in $d=3$) to $T^2 = +1$ (in $d=4$), which is also a potentially serious change where disorder physics is concerned. (For a more in-depth discussion of issues with  Abrikosov's continuation, see Ref.~\cite{rahul-sid}). 

Thus we employ instead the RG scheme developed by Moon {\it et al} \cite{MoonXuKimBalents}. In this RG scheme, the radial momentum integrals are performed with respect to a $d=4-\epsilon$ dimensional measure $\int \frac{p^{3-\epsilon} dp}{(2\pi)^{4-\epsilon}}$, but the $\Gamma$ matrix structure is as in $d=3$, and the angular integrals are performed only over the three-dimensional sphere parametrized by the polar and azimuthal angles $(\theta, \,\varphi)$. Nevertheless, the overall angular integral of an isotropic function $\int_{\hat{\Omega}}\cdot 1$ is taken to be $2 \,\pi^2$ (as is appropriate for the total solid angle in  $d=4$), and the angular integrals are normalized accordingly. Therefore, the angular integrations are performed with respect to the measure 
\be\label{moonmeasure}
\int dS\, (\ldots) \equiv \frac{\pi}{2} \int_0^{\pi} d \theta \int_0^{2\pi} d \varphi\, \sin \theta \, (\ldots)\,,\ee
 where the $\pi/2$ is inserted for the sake of normalization. For more details on this renormalization scheme, see \cite{MoonXuKimBalents, rahul-sid}. 
For performing the angular integrals, we will use the notation 
\be
 \hat{d}_a(\vec{k}) =\frac{\tilde d_{a}(\vec{k}) } {k^2} \,,
\ee such that
\be
\label{ang}
\int dS \, \hat{d}_{a}(\vec{k}) = 0\,, \quad
\int dS\, \hat{d}_{a}(\vec{k})\, \hat{d}_{b}(\vec{k}) =\frac{ 2\,\pi^2\, \delta_{ab}}  {N} \,.
\ee
We note that $|\dvec(\vec{k})|^2 =\frac{ k^{4}}{4\,m^2}$.

 %%%%%%%%%%%%%%%%%%%%%%%%%%%%%%%%%%%%%%%%%%%%%%
\begin{figure}
\subfloat[]{\includegraphics[width = 0.3 \columnwidth]{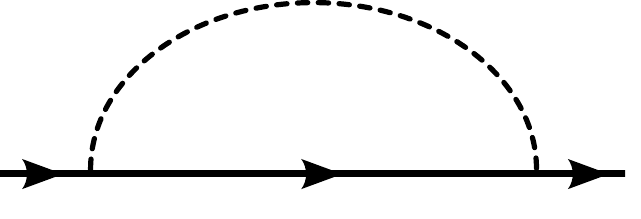}} \hspace{1 cm}
\subfloat[]{\includegraphics[width = 0.3\columnwidth]{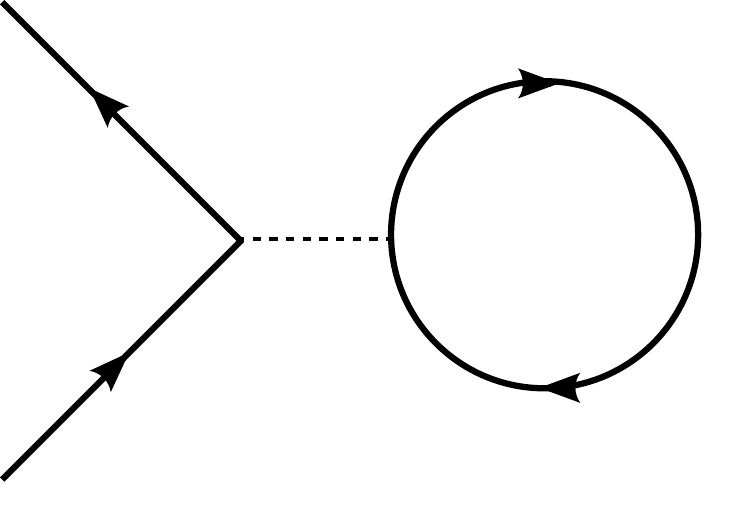}}
\caption{\label{scalardisorderonlygf} These two diagrams determine the $O(\epsilon)$ correction to the Green's function with each solid line representing the bare Green's function. A dashed line may represent either a disorder or the Coulomb interaction. If the dashed line represents disorder, then it connects two fermion lines at the same point in real space, but they may have different time and replica indices. For disorder interaction, the diagram (b) is then proportional to the number of replica flavours $n$, and vanishes upon taking the replica limit $n \rightarrow 0$.  If the dashed line represents the Coulomb interaction, then it connects two fermions with the same time index and same replica index, but with different spatial positions. }
\end{figure}

%%%%%%
\begin{figure}
\subfloat[]{ \includegraphics[width = 0.35 \columnwidth]{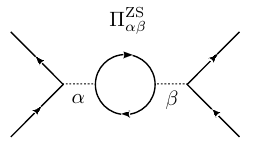}}  \hspace{1 cm}
\subfloat[]{ \includegraphics[width = 0.35 \columnwidth]{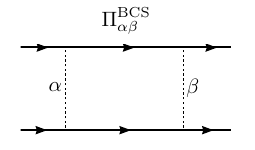}}
\\
\subfloat[]{\includegraphics[width = 0.35 \columnwidth]{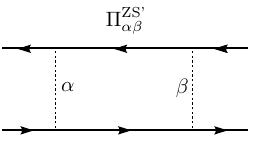}}  \hspace{1 cm}
\subfloat[]{ \includegraphics[width = 0.35 \columnwidth]{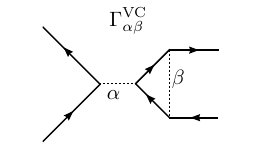}}
\caption{\label{scalardisordervertices} These diagrams determine the $O(\epsilon)$ correction to the four-fermion vertices (either disorder or interaction) and are denoted by ZS, BCS, ZS$'$ and VC, respectively, following the convention of Ref.~\cite{Shankar}. Each solid line represents the bare Green function. A dashed line may represent either a disorder or the Coulomb interaction. Note that unlike the ZS, BCS, and ZS$'$ diagrams,  the VC diagrams are generically not symmetric under interchange of indices, \textit{i.e.} $\Gamma^{\text{VC}}_{\alpha\beta}\neq\Gamma^{\text{VC}}_{\beta\alpha}$.  } 
\end{figure}

We will consider RG flows by considering the one-loop corrections coming from the diagrams shown in Figs~\ref{scalardisorderonlygf}~and~\ref{scalardisordervertices}. We will employ the momentum-shell RG, and
consider the RG flow generated by changing
$\Lambda_{\text{UV}} / \Lambda_{\text{IR}}$ as $e^{-l}$, where $\Lambda_\text{UV}\,(\Lambda_\text{IR})$ is the UV (IR) cut-off for the energy/momentum integrals and $l$ is the RG flow parameter. We will use the one-loop $\beta$ functions, that dictate the  flow of the parameters with increasing $l$. Furthermore, in our RG scheme we will hold $m$ fixed, \text{i.e.} $m$ does not flow such that
\begin{align}
\frac{d m} {dl}=0 \,.%\Rightarrow (z-2)\,m=0
\label{rgm}
\end{align}
This is simply a choice. Any scale dependence of $m$ is absorbed into a scale dependence of $\psi$, such that $\psi$ acquires an anomalous dimension, but $m$ remains fixed. 

%%%%%%%%%%%%%%%%%%%%%%%%%%%%%%%%%%%%%%%%
\section{Including time-reversal symmetry breaking disorder}
\label{sec:ph-sym}
In this section we incorporate time-reversal symmetry breaking disorder, while continuing to assume equal masses for the conduction and valence bands. In the absence of the $\frac{k^2}{2\,m'}$ term, the bare Green's function for each fermionic flavour is given by:
\begin{equation}
G_0(\omega, \vec{k}) = \frac{1} {- i\, \omega + \dvec(\vec{k}) \cdot{\boldsymbol{\Gamma}}} =  \frac{i\, \omega + \dvec(\vec{k}) \cdot{\boldsymbol{\Gamma}}}   {\omega^2 +|\dvec(\vec{k})|^2} \,,
\label{baregf}
\end{equation}
here $|d(\vec{k})|^2 = (\frac{k^2}{2m})^2$. On occasions, to lighten the notation, we will use $d_{\vec{k}}$ to denote $d(\vec{k})$. 
The Abrikosov fixed point for the clean system is given by:
\begin{align}
u=u^*\equiv \frac{15}{2\left( 4 +15  \, N_f\right )},
\quad \text{ where } u = \frac{m\,e^2}{8 \,\pi^2\, c}\,.
\end{align}

All diagrams not involving $W_2$ lines were computed in \cite{rahul-sid}. Our immediate challenge is to augment that analysis with time-reversal symmetry breaking disorder, $W_2\neq 0$
\subsection{Addition of the $W_2$ vertex to the non-interacting problem}
%%%%%%%%%%
The loop corrections to the disorder lines themselves come from the fully connected contractions of
\begin{align}
\label{disordercontraction}
 \delta S 
  =&  \frac12 \int  d\tau\, d\tau'\,  d\tau''\, d\tau'''\,d^d x\,  d^d x' 
   \sum \limits_{i,j,k,l, \xi}
 \Big [ W_0^2\, ({ \psi_i^\xi}^{\dag}  {\psi_i^\xi})_{x}^{\tau}\, ({ \psi_j^\xi}^{\dag} \psi_j)_{x}^{\tau'} \,
( {\psi_k^\xi}^{\dag} \,\psi_k^\xi )_{x'}^{\tau''} \,( {\psi^\xi_l}^{\dag}\, \psi_l^\xi )_{x'}^{\tau'''} \nn 
& \hspace{ 5.4 cm}
+ W_1^2\, ({ \psi_i^\xi}^{\dag} \,\Gamma^i_{a} \, {\psi_i^\xi})_{x}^{\tau}\, ({ \psi_j^\xi}^{\dag} \,\Gamma^j_{a} \,\psi_j^\xi )_{x}^{\tau'} \,
( {\psi_k^\xi}^{\dag}\, \Gamma^k_{b}\, \psi_k^\xi )_{x'}^{\tau''}\, ( {\psi^\xi_l  }^{\dag}\,\Gamma^l_{b}\, { \psi_l^\xi})_{x'}^{\tau'''}
\nn & \hspace{ 5.4 cm}
+  2\, W_0 \,W_1 \, ({ \psi_i^\xi}^{\dag}  {\psi_i^\xi})_{x}^{\tau} \,({ \psi_j^\xi}^{\dag} \,\psi_j^\xi  )_{x}^{\tau'} \,
(  {\psi_k^\xi}^{\dag} \,\Gamma^k_{b} \,\psi_k^\xi )_{x'}^{\tau''} \,( {\psi^\xi_l }^{\dag}\, \Gamma^l_{b}\,{ \psi_l^\xi})_{x'}^{\tau'''} \nn
&\hspace{ 5.4 cm}
+ W_2^2\, ({ \psi_i^\xi}^{\dag}\, \Gamma^i_{ab} \, {\psi_i^\xi})_{x}^{\tau}\, ({ \psi_j^\xi}^{\dag} \,\Gamma^j_{ab} \,\psi_j)_{x}^{\tau'} \,
({ \psi_k^\xi}^{\dag} \,\Gamma^k_{cd}\, \psi_k)_{x'}^{\tau''}\, ({\psi_l^\xi}^{\dag}\, \Gamma^l_{cd}\, { \psi_l^\xi})_{x'}^{\tau'''}
%%%%%%%%%%%%%
\nn & \hspace{ 5.4 cm}
+  2\, W_0 \,W_2 \, ({ \psi_i^\xi}^{\dag}\,  {\psi_i^\xi})_{x}^{\tau} \,({ \psi_j^\xi}^{\dag} \psi_j^\xi )_{x}^{\tau'}\,\,
( {\psi^\xi_k}^{\dag}\, \Gamma^k_{cd}\, \psi_k^\xi )_{x'}^{\tau''} \,( {\psi_l^\xi}^{\dag} \,\Gamma^l_{cd}\,  { \psi_l^\xi})_{x'}^{\tau'''} \nn
& \hspace{ 5.4 cm}
+  2\, W_1 \,W_2 \, ({ \psi_i^\xi}^{\dag}\,\Gamma^i_{a} \, {\psi_i^\xi})_{x}^{\tau} \,({ \psi_j^\xi}^{\dag} \,\Gamma^j_{a}\, \psi_j^\xi)_{x}^{\tau'}\,
({\psi^\xi_k  }^{\dag}\, \Gamma^k_{cd} \,\psi_k)_{x'}^{\tau''} \,({\psi^\xi_l}^{\dag}\, \Gamma^l_{cd}   \, { \psi_l^\xi} )_{x'}^{\tau'''}
\Big  ]   \,,
\end{align}
where repeated $\Gamma$-matrix indices are as usual summed over, and {we have kept track of the replica indices on $\Gamma$ matrices}. We note that we must incorporate $a\neq b$ ($c\neq d$) for the sum over $\Gamma_{ab}$ ($\Gamma_{cd}$) matrices as we should consider only the independent terms.

\subsubsection{Fermion self-energy}
The correction to the one-loop fermion self-energy from tensor disorder is given by:
\begin{align}
\Sigma_{W_2} (\omega,\mathbf k)&=2\, W_2 \int  \frac{d^d p} {(2\, \pi)^d} \Gamma_{ab}  \, G (\omega,\mathbf p)\,  \Gamma_{ab}  
= \frac{i   \,\omega \,m^2 \, W_2     \, N \left ( N-1\right) \ln \left (  \frac {\Lambda_{UV} }  {\Lambda_{IR} }\right )} { 2 \, \pi^2}  \,.
\end{align}

In the presence of the tensor disorder, the dynamical exponent is thus modified to:
\begin{align}
\label{dynexp}
z&=2+ \frac{m^2} {\pi^2} \Big [ W_0 + N W_1 + \frac{N \left ( N-1\right) W_2}{2} \Big]
= 2+\frac{ \lambda_0 +  N \lambda_1 + \frac{N \left ( N-1\right) \lambda_2 } {2} } {2}\,,
\end{align}
where
\begin{align}
\lambda_\alpha = \frac{\pi^2\, W_\alpha}{2\,m^2} \quad \text{for } \alpha=0,1,2\,.
\end{align}

\subsubsection{ZS diagrams}

These are zero.

\subsubsection{VC diagrams}
A VC with two tensor ($W_2$) lines, emerging from 8 distinct contractions, and after setting the external frequency $\omega = 0$, leads to:
\begin{align}
 \frac{ \Gamma_{22}^{ \text{VC} } }   {(2\, m)^2} 
&= 4 \,W_2^2 \, \Gamma^i_{ab} \int \frac{d^dk} {(2\pi)^d}
\frac{ \Gamma_{cd}^j \left (\hat{\mathbf{d_k}} \cdot \mathbf{ \Gamma}^j \right )   \Gamma^j_{ab}   \left (\hat{\mathbf{d_k}} \cdot \mathbf{ \Gamma}^j \right )   \Gamma^j_{cd}   } 
 {k^4}\nn
& = \frac{W_2^2  \, \Gamma^i_{ab} \,   \Gamma^j_{cd}\,   \Gamma^j_{f}\,  \Gamma^j_{ab}\,  \Gamma^j_{f}\,   \Gamma^j_{cd}} 
 {2N \pi^2} \ln \left (  \frac {\Lambda_{UV} }  {\Lambda_{IR} }\right )\nn
&= - \frac{W_2^2  \left (  N-4 \right ) \Gamma^i_{ab} \,   \Gamma^j_{cd}\,  \Gamma^j_{ab} \Gamma^j_{cd}\, l} 
 {2N \pi^2} \nn
&=  -\frac{W_2^2  \left (  N-4 \right ) \left (N^2-9N + 16 \right )
\Gamma^i_{ab} \,  \Gamma^j_{ab}\,l  } 
 { 4 N \pi^2}\,,
\end{align}
using Eqs.~(\ref{rel3}) and (\ref{rel4}).
The correction is thus given by $\delta \lambda_2 = \frac{ \left (  N-4 \right ) \left (N^2-9N +16 \right ) }{2 N}\, \lambda_2^2 \, l$ . 

\begin{align}
\frac{ \Gamma_{02}^{ \text{VC} } }   {(2\, m)^2}
&= 4\,W_0 \, W_2   \int \frac{d^dk} {(2\pi)^d}
\frac{ \Gamma_{cd}^j \left (\hat{\mathbf{d_k}} \cdot \mathbf{ \Gamma}^j \right )    \left (\hat{\mathbf{d_k}} \cdot \mathbf{ \Gamma}^j \right )   \Gamma^j_{cd}   } 
 {k^4}\nn
& = \frac{W_0 \, W_2 \, \Gamma_{cd}^j\,\Gamma_{f}^j\, \Gamma_{f}^j\, \Gamma_{cd}^j} {2 N \pi^2}  \ln \left (  \frac {\Lambda_{UV} }  {\Lambda_{IR} }\right )\nn
& = - \frac{W_0 \, W_2\, N \left ( N-1 \right ) l} {4  \,\pi^2} \,.
\label{gammma02}
\end{align}
This gives the correction $\delta \lambda_0 =\frac{ N \left ( N-1 \right ) }{2}\lambda_0\, \lambda_2 \, l$.
%%%%%%%%%%%%%%%%%%

\begin{align}
\frac{ \Gamma_{2 0}^{ \text{VC} } }   {(2\, m)^2}
&= 4W_2 \, W_0\, \Gamma^i_{ab}   \int \frac{d^dk} {(2\pi)^d}
\frac{  \left (\hat{\mathbf{d_k}} \cdot \mathbf{ \Gamma}^j \right )  \Gamma^j_{ab}    \left (\hat{\mathbf{d_k}} \cdot \mathbf{ \Gamma}^j \right )    } 
 {k^4}\nn
& = \frac{ W_2 \, W_0\, \Gamma^i_{ab} \,\Gamma_{f}^j    \, \Gamma_{a b}^j\, \Gamma_{f}^j} {2 N \pi^2}  \ln \left (  \frac {\Lambda_{UV} }  {\Lambda_{IR} }\right )\nn
& =- \frac{ W_0 \, W_2 \left ( N-4 \right ) \Gamma^i_{ab} \,  \Gamma_{a b}^j \, l} {2 N \pi^2}  \,,
\end{align}
%%%%%%%%%%%%%%%%%%
using Eq.~(\ref{rel4}).
The contribution from this term is therefore $\delta \lambda_2 = \frac{  N-4 }{N}\, \lambda_0 \,\lambda_2 \, l $.

\begin{align}
 \frac{ \Gamma_{12}^{ \text{VC} } }   {(2\, m)^2} 
&= 4 W_1\, W_2 \, \Gamma^i_{a} \int \frac{d^dk} {(2\pi)^d}
\frac{ \Gamma_{cd}^j \left (\hat{\mathbf{d_k}} \cdot \mathbf{ \Gamma}^j \right )   \Gamma^j_{a}   \left (\hat{\mathbf{d_k}} \cdot \mathbf{ \Gamma}^j \right )   \Gamma^j_{cd}   } 
 {k^4}\nn
%%%%%%%%%
& = \frac{  W_1 \,W_2  \, \Gamma^i_{a} \,   \Gamma^j_{cd}\,   \Gamma^j_{f}\,  \Gamma^j_{a}\,  \Gamma^j_{f}\,   \Gamma^j_{cd}} 
 {2N \pi^2} \ln \left (  \frac {\Lambda_{UV} }  {\Lambda_{IR} }\right )\nn
& = \frac{  W_1 \,W_2  \left ( 2-N \right ) \Gamma^i_{a} \,   \Gamma^j_{cd}\,  \Gamma^j_{a}\,  \Gamma^j_{cd}} 
 {2N \pi^2} \ln \left (  \frac {\Lambda_{UV} }  {\Lambda_{IR} }\right )\nn
%%%%%%%%%%%%%
&= \frac{  W_1\, W_2 \,\left ( N-1 \right )  \left ( N-2 \right )  \left (N-4 \right )\Gamma^i_{a}  \, \Gamma^j_{a}\,l } 
 {4  N \pi^2} \,,
\end{align}
%%%%%%%%%%%%%%%%
where we have used Eq.~(\ref{rel3}).
This gives the correction $\delta \lambda_1 =- \frac{ \left ( N-1 \right )  \left ( N-2 \right )  \left (N-4 \right ) }{ 2 N}\, \lambda_1\, \lambda_2 \, l$ .

\begin{align}
 \frac{ \Gamma_{2 1}^{ \text{VC} } }   {(2\, m)^2} 
& = 4 W_2\, W_1 \, \Gamma^i_{a b} \int \frac{d^dk} {(2\pi)^d}
\frac{ \Gamma_{c}^j \left (\hat{\mathbf{d_k}} \cdot \mathbf{ \Gamma}^j \right )   \Gamma^j_{a b}   \left (\hat{\mathbf{d_k}} \cdot \mathbf{ \Gamma}^j \right )   \Gamma^j_{c}   } 
 {k^4}\nn
& = \frac{  W_2 \,W_1  \, \Gamma^i_{a b} \,   \Gamma^j_{c}\,   \Gamma^j_{f}\,  \Gamma^j_{a b}\,  \Gamma^j_{f}\,   \Gamma^j_{c}\, \ln \left (  \frac {\Lambda_{UV} }  {\Lambda_{IR} }\right )} 
 {2N \pi^2} \nn
&=- \frac{  W_1\, W_2 \left( N-4 \right )^2
 \Gamma^i_{ab}  \, \Gamma^j_{a b}\,l } 
 {2N \pi^2} \,,
\end{align}
%%%%%%%%%%%%%%%%
using Eq.~(\ref{rel4}).
The correction from this term is $\delta \lambda_2 = \frac{  \left(N-4 \right )^2}{N}\, \lambda_1 \,\lambda_2 \, l $.

%%%%%%%%%%%%%%%%%%%%%%%%%%%%%%%%%
\subsubsection{BCS and ZS$\,'$ diagrams}

\begin{align}
\frac{\Pi^{\text{BCS}}_{22}}{(2\,m)^2}
& =2\, W_2^2 \int \frac{d^d k}  {(2 \pi)^d}
\frac{ \Big[  \Gamma^i_{ab} \left (\hat{\mathbf{d_k}} \cdot \mathbf{ \Gamma}^i \right )  \Gamma^i_{ c d}\Big ] \,
 \Big[  \Gamma^j_{ c d} \left (\hat{\mathbf{d_k}} \cdot \mathbf{ \Gamma}^j \right )  \Gamma^j_{ ab }\Big ]}
{k^4}\,,\nn
%%%%%%%%%%%%%%
\frac{\Pi^{\text{ZS}'}_{22}}{(2 \,m)^2}
 & =2 \,W_2^2 \int \frac{d^d k}  {(2 \pi)^d}
\frac{ \Big[  \Gamma^i_{ab} \left (\hat{\mathbf{d_k}} \cdot \mathbf{ \Gamma}^i \right )  \Gamma^i_{ c d}\Big ] \,
 \Big[  \Gamma^j_{ a b} \left (\hat{\mathbf{d_k}} \cdot \mathbf{ \Gamma}^j \right )  \Gamma^j_{ cd }\Big ]}
{k^4} \,.
\end{align}
%%%%%%%%%%%%%%%%%%%%%%%
Adding these together, we get:
\begin{align}
 \frac{\Pi^{\text{BCS}}_{22}}{(2\, m)^2} + \frac{\Pi^{\text{ZS}'}_{22}}{(2 \,m)^2}
=\frac{W_2 ^2\, \ln \left (  \frac {\Lambda_{UV} }  {\Lambda_{IR} }\right )
 \Gamma^i_{ab}\,  \Gamma^i_{e} \,  \Gamma^i_{ c d} 
\left (  \Gamma^j_{c d} \,  \Gamma^j_{e} \,  \Gamma^j_{ab}  
+  \Gamma^j_{ab}  \,  \Gamma^j_{e} \,  \Gamma^j_{c d}\right ) }  {4 N \pi^2 } 
 = -\frac{W_2 ^2 \, l   \left ( 3 
 + \frac{17} {5} \sum \limits_{a} \Gamma_a^i\,\Gamma_a^j \right)}  {  \pi^2 } \,,
\end{align}
where we have used Eq.~(\ref{rel53}).
This corrects both the scalar and vector disorder terms by
$\delta \lambda_0 =  6  \, \lambda_2^2 \, l $
and $\delta \lambda_1 = { \frac{ 34 } {5} } \, \lambda_2^2 \, l $, respectively.

%%%%%%%%%%%%%
\begin{align}
\frac{\Pi^{\text{BCS}}_{02}}{(2\, m)^2} + \frac{\Pi^{\text{ZS}'}_{02}}{(2\,m)^2}
 &= 4 W_0 \,W_2   \int \frac{d^d k  \, k^{-4} }  {(2 \pi)^d}
\sum \limits_{a< b}   \Big[  \Gamma^i_{ab} \left (\hat{\mathbf{d_k}} \cdot \mathbf{ \Gamma}^i \right )   \Big ] 
 \Big[  \Gamma^j_{ ab} \left (\hat{\mathbf{d_k}} \cdot \mathbf{ \Gamma}^j \right )
 + \left (\hat{\mathbf{d_k}} \cdot \mathbf{ \Gamma}^j \right ) \Gamma^j_{ ab }\Big ]
 \nn
 %%%%%%%%%%%%%%%%%%%%%%55
&=\frac{W_0\, W_2 }  { 2 N \pi^2 } \ln \left (  \frac {\Lambda_{UV} }  {\Lambda_{IR} }\right )
\sum \limits_{a \neq b,\, f}
 \Gamma^i_{ab}\,  \Gamma^i_{e}  
\left (  \Gamma^j_{ab} \,  \Gamma^j_{e} +   \Gamma^j_{e} \,  \Gamma^j_{a b}\right )\nn
&
=-\frac{ 3\,W_0\, W_2 \,l}  { 10\, \pi^2 } \,
\sum \limits_{a < b} \Gamma_{ab}^i \, \Gamma_{ab}^j  \,,
\end{align}
where we have used Eq.~(\ref{rel51}).
%{\color{red}This one}
This corrects the tensor disorder term by 
$\delta \lambda_2 =\frac{3} {5 }\, \lambda_0\, \lambda_2  \, l$.

\begin{align}
& \frac{\Pi^{\text{BCS}}_{12}}{(2\, m)^2} + \frac{\Pi^{\text{ZS}'}_{12}}{(2\, m)^2} 
%%%%%%%%%%%%%%
= 4 W_1 \,W_2   \int \frac{d^d k \, k^{-4}}  {(2 \pi)^d} \sum \limits_{a< b,\,c}
\Big[  \Gamma^i_{ab} \left (\hat{\mathbf{d_k}} \cdot \mathbf{ \Gamma}^i \right )   \Gamma^i_{ c}   \Big ] 
\times \Big[  \Gamma^j_{ ab} \left (\hat{\mathbf{d_k}} \cdot \mathbf{ \Gamma}^j \right )   \Gamma^j_{ c} 
+  \Gamma^j_{ c}   \left (\hat{\mathbf{d_k}} \cdot \mathbf{ \Gamma}^j \right ) \Gamma^j_{ ab }\Big ] \nn
%%%%%%%%%%%%%
&=\frac{W_1 \,W_2\,\ln \left (  \frac {\Lambda_{UV} }  {\Lambda_{IR} }\right )
 \sum \limits_{a \neq  b,\,c, \, f} \Gamma^i_{ab}\,  \Gamma^i_{f} \,  \Gamma^i_{ c } 
\left (  \Gamma^j_{c } \,  \Gamma^j_{f} \,  \Gamma^j_{ab}  +  \Gamma^j_{ab}  \,  \Gamma^j_{f} \,  \Gamma^j_{c }\right ) }  { 2 N \pi^2 }
=
- \frac{ 17\,W_1 \,W_2\,l }  { 5\, \pi^2 } \,\sum \limits_{a< b} \Gamma_{ab}^i\, \Gamma_{ab}^j
\,,
\label{gammaal}
\end{align}
where we have used Eq.~(\ref{rel52}).
%{\color{red}This one}
This corrects the tensor disorder term by 
$\delta \lambda_2 =\frac{34} { 5 }\, \lambda_1\, \lambda_2  \, l$.

%%%%%%%%%%%%%%%%%%%%%%%%%%
\subsubsection{RG equations}

The tree-level scaling dimension of the disorder term is $(2\, z - d)=\varepsilon  +\lambda_0 +N\lambda_1 +\frac{N\left( N-1\right )\lambda_2} {2}$, where $\varepsilon = 4-d$. Using Tables~\ref{VC}~and~\ref{BCS}, the RG equations for the disorder couplings are thus given by:
%%%%%%%%%%%%%%%

\begin{table}
\begin{tabular}{|c|c|c|c||c|}
\hline
Coupling  & $ \lambda_0$ & $ \lambda_1 $ & $\lambda_2$ & $u$ \tabularnewline
\hline
$ \lambda_0$ & $\delta \lambda_0 = \lambda_0^2  \, l $ & 
% THIS ONE
$\delta \lambda_0=  N\, \lambda_0\, \lambda_1\,l$
&   
$ \delta \lambda_0 =\frac{ N \left ( N-1 \right )\lambda_0\, \lambda_2 \, l }{2}$
 & 0 \tabularnewline
\hline
%%%%%%%%%%%%%%%%%%%%%%%%%%%%%%%%
$ \lambda_1$ &
%THIS ONE 
$\delta \lambda_1 = -   \frac{\left (N-2\right ) \lambda_0 \, \lambda_1\, l }{N } $ 
&$\delta \lambda_1 =  \frac{(N-2)^2 \, \lambda_1^2\, l}{ N}$& 
\makecell{$ \delta \lambda_1  = - \frac{ \left ( N-1 \right )  \left ( N-2 \right )  \left (N-4 \right )\lambda_1\, \lambda_2 \, l }{ 2 N}  $}
& \makecell{$ \delta  \lambda_1  =    \frac{2 \left (N-1 \right )  \lambda_1\, u  \,l}{N } $}
 \tabularnewline
\hline
%%%%%%%%%%%%%%%%%%%%%%%%%%%%%%%%%%%%%%%
$ \lambda_2$ &  
\makecell{$\delta \lambda_2 $\\
$= \frac{ \left (  N-4 \right ) \lambda_0 \,\lambda_2 \, l }{N} $}
& \makecell{$\delta \lambda_2  = \frac{  \left(N-4 \right )^2  \lambda_1 \,\lambda_2 \, l }{N}$}&
\makecell{$ \delta \lambda_2 = \frac{ \left (  N-4 \right ) \left (N^2-9N + 16 \right ) \lambda_2^2 \, l  }{2 N}$}
& $d\lambda_2=\frac{4 \,\lambda_2\, u  \, l } {N}  $ \tabularnewline
\hhline{|=|=|=|=||=|}
$u$ & $\delta u = \lambda_0\, u \, l $ & 
$ \delta u =  N\, \lambda_1 \,u \, l$ & 
$ \delta u= \frac{  N\left ( N-1 \right )  \lambda_2 \,u\, l} { 2   } $
 &0\tabularnewline
\hline
\end{tabular}
\caption{Contributions to the $\beta$-functions from the VC diagrams without the $\frac{k^2}{2m'}$ term. Here, $\lambda_\alpha = \frac{2\,m^2\,W_\alpha}{\pi^2}$, $u = \frac{m\,e^2}{8 \,\pi^2\, c}$, and $l$ is the RG flow parameter. Terms not involving $W_2$ are taken from Ref.~\cite{rahul-sid}. \label{VC}} 
\end{table}

%%%%%%%%%%%%%%%%%%%%%%%%%%%%%%
 \begin{table}
\begin{tabular}{|c|c|c|c||c|}
\hline
Coupling  & $ \lambda_0$ & $ \lambda_1 $ & $\lambda_2 $ & $u$ \tabularnewline
\hline
$ \lambda_0$ &$\delta \lambda_1 = \frac{1}{N}\, \lambda_0^2\, l$
& $\delta \lambda_0 = 2\, \lambda_0\, \lambda_1\,l$
&
$\delta \lambda_2 =\frac{3} {5}\, \lambda_0\, \lambda_2  \, l$
& $ 0$
 \tabularnewline
\hline
$\lambda_1$ &
\makecell{included in $(\lambda_0, \,\lambda_1)$ cell}
&$ \delta \lambda_1 =   \frac{3 N-2}{N}\,\lambda_1^2\, l$ & 
 $\delta \lambda_2 =\frac{34}{ 5 }\, \lambda_1\, \lambda_2  \, l$
&$0$ \tabularnewline
\hline
%%%
$\lambda_2 $ & \makecell{included in $(\lambda_0, \,\lambda_2)$ cell}
& \makecell{included in $(\lambda_1, \,\lambda_2)$ cell} &
%%%%%%%%%%%%%%%%%%%
\makecell{$\delta \lambda_0 = 6 \, \lambda_2^2 \, l $,\\  
$\delta \lambda_1 = { \frac{ 34 } {5} } \, \lambda_2^2 \, l $}
%%%%%%%%%%%%%%%%%%%%%%%%5
& $0$ \tabularnewline
\hhline{|=|=|=|=||=|}
$u$ &  $0$ & $0$ &     $0$ &  $0$\tabularnewline
\hline
\end{tabular}
\caption{Sum of the contributions to the $\beta$-functions from the BCS and ZS$'$ diagrams without the $\frac{k^2}{2m'}$ term, using the same conventions as Table~\ref{VC}. \label{BCS}}
\end{table}

%%%%%%%%%%%%%%%%%%%%%%%%%%%%%%%%%%%%%%%
\begin{align}
\label{rgflow}
\frac{d\lambda_0} {dl} = & \left[\varepsilon+2 \,\lambda_0 +2\left (N+1\right ) \lambda_1 + N \left( N-1\right ) \lambda_2\right ]  \lambda_0  + 6\,\lambda_2^2 \, ,\\
%%%
\frac{d\lambda_1 } {dl} = & \left [ \epsilon 
+\left( 2\,N - 1 \right) \lambda_1 + \left( 3\,N-7  \right) \lambda_{2}
+\frac{2 \left (\lambda_{0} +\lambda_{1}+2 \,\lambda_{2}\right )} {N}
\right  ]  \lambda_1
+\frac{\lambda_0^2} {N} +\frac{ 34\,\lambda_2^2} {5} \,,\\
%%%%%
\frac{d\lambda_2  } {dl} = & 
 \left[ \epsilon +
 \frac{ 13\, \lambda _0 - 6 \, \lambda _1 }  {5}
+2 \,N\,\lambda _1
+ \frac{4 \left(  4 \,\lambda _1 - \lambda _0\right)} {N} \right ] \lambda _2
+  \left(N^2-7\, N-\frac{32}{N}+26 \right) \lambda _2^2 \,.
\end{align}
%%%%%%%%%%%%%%%%%%%%%%%%%
Analysis of these equations is deferred to Sec.~\ref{secrg}~and~\ref{strong}.

%%%%%%%%%%%%%%%%%%%%%%%%%%%%%%%
\subsection{Addition of the $W_2$ vertex in the presence of Coulomb interactions} 
\label{couldis}

\subsubsection{ZS diagram}
The ZS diagram with one Coulomb line and one $W_2$ line attached vanishes upon tracing over spinor indices.

\begin{table}
\begin{tabular}{|c|c|c|c||c|}
\hline
Coupling  & $ \lambda_0$ & $ \lambda_1 $ & $ \lambda_2$ & $u$ \tabularnewline
\hline
$ \lambda_0$ &0& $0$& 0& $ \delta \lambda_0 = - 4 N_f\,  \lambda_0\, u\,l $ \tabularnewline
\hline
$\lambda_1$ &0&0 & 0&0 \tabularnewline
\hline
 $ \lambda_2$ &0&0&0&0 \tabularnewline
\hhline{|=|=|=|=||=|}
$u$ &0&0 &0 & $\delta  u =  - 2 N_f\, u^2\,l $ \tabularnewline
\hline
\end{tabular}
\caption{Contributions to the $\beta$-functions from the ZS diagram, with the same notation as in Tables~\ref{VC} and~\ref{BCS}. The results are valid for the cases with and without the $\frac{k^2}{2m'}$ term.
\label{ZS}}
\end{table}

\subsubsection{VC diagrams}

The Coulomb correction to the $W_2$ vertex takes the form:
%%%%%%%%%%%%%%%%%%%
\begin{align}
 \Gamma^{\text{VC}}_{2c} 
&=-\frac{ 2 \, W_2\, e^2\,  \Gamma^i_{ ab}  } {  c} \int \frac{d\omega \, d^{d } p}  {(2\, \pi)^{d+1} }
\frac{ \Big[ i \, \omega +   \left ( {\mathbf{d_p}} \cdot \mathbf{ \Gamma}^j \right )   \Big ] \,   \Gamma^j_{ ab} 
 \Big[ i \, \omega +   \left ( {\mathbf{d_p}} \cdot \mathbf{ \Gamma}^j \right )   \Big ]}
{ p^2  \left(  \omega^2  + \mathbf{d_p}^2   \right)^2   } \nn
&=-\frac{ 2 \, W_2\, e^2\,  \Gamma^i_{ ab}  \times 2\, \pi^2 } {  c} \int \frac{d\omega \, d p\, p^3}  {(2 \, \pi)^{d+1} }
\frac{ -\omega^2 \,\Gamma^j_{ab} +\frac{p^4}  {2\,m^2 N}  \,\Gamma^j_{ f}\, \Gamma^j_{ ab} \, \Gamma^j_{ f } }
{ p^2  \left(  \omega^2  +\frac{p^4}  { 4 \,m^2 }  \right)^2   } \nn
&= \frac{ 2 \, \pi \,W_2\, e^2\,  \Gamma^i_{ ab}  } {  c} \int \frac{d\omega \, d  p \, p^3}  {(2 \,\pi)^{d} }
\frac{  (\omega^2  - \frac{\left (N-4 \right)   p^4}  { 4 \,m^2 N}   }
{ p^2  \left(  \omega^2  + \frac{p^4}  { 4\, m^2 }   \right)^2   }\, \Gamma^j_{ab} \nn
& =  \frac{    W_2\, e^2\,  \Gamma^i_{ ab} \,\Gamma^j_{ab}   } { 2 \,\pi^2\,  c} 
\int 
\frac{  d  p \,m  }  
{   N\,p      }
%%%%%%%
 = \frac{    W_2\,m\, e^2 \,   \Gamma^i_{ ab} \,\Gamma^j_{ab}  \ln \left (  \frac {\Lambda_{UV} }  {\Lambda_{IR} }\right )  } { 2 \, \pi^2  \,  c\, N} 
 =- \frac{    W_2\,m\, e^2\,  \Gamma^i_{ ab} \,\Gamma^j_{ab} \, l } { 2\, \pi^2  \,  c\, N} \,.
\label{gamma2c}
\end{align}
Here we have used Eqs.~(\ref{ang}) and(\ref{rel3}). The above gives a correction $d\lambda_2=\frac{4\,\lambda_2\, u  \, l } {N}$.

The tensor disorder correction to the Coulomb vertex is given by:
\begin{align}
\Gamma^{\text{VC}}_{c2}=- \frac{e^2 } {2\,c\,q^2 \, W_0}\times  \Gamma^{\text{VC}}_{02}
 =- \frac{ m^2  e^2 \,W_2 \,N \left ( N-1 \right ) l} { 2 \,c\, q^2 \, \pi^2} \,,
\end{align}
using Eq.~(\ref{gammma02}).
This gives the correction as $\delta u = \frac{ m^2 \,W_2\, u \,N \left ( N-1 \right ) l} {  \pi^2}= \frac{ \lambda_2 \,u\, N\left ( N-1 \right ) l} { 2   } $, where an additional minus sign has to be taken into account.

\subsubsection{BCS and ZS$\,'$ diagrams}

These make a vanishing contribution, for reasons discussed in Ref.~\cite{rahul-sid}. 

%%%%%%%%%%%%%%%%%%%%%%%%%%%%%%%%
\subsection{RG equations}
\label{secrg}

In the presence of Coulomb interactions, the dynamical critical exponent is given by:
\begin{align}
\label{dynexpcol}
z
= 2+\frac{ \lambda_0 +  N \lambda_1 + \frac{N \left ( N-1\right) \lambda_2}{2} } {2} 
- \frac{8\, u}{15} \,,
\end{align}
where $u=\frac{m\, e^2}{8\, \pi^2\, c}$.
Now, the tree-level scaling dimension of the disorder term becomes
$$(2\, z - d)=\varepsilon  +\lambda_0 +N\lambda_1 +\frac{N\left( N-1\right )\lambda_2} {2}
- \frac{ 16\, u}{15} \,.$$

Using Tables~\ref{VC},~\ref{BCS}~and~\ref{ZS}, the full set of the RG equations for the disorder couplings as well as $u$ is given by:
%%%
\begin{align}
\label{l0}
\frac{d\lambda_0} {dl} = & \left[ \epsilon
-\frac{4\,u}{15}  \left(15 \,N_f+4\right)+ 2\, \lambda _0
+2 \left (N+1 \right) \lambda _1 +N \left (N-1 \right ) \lambda _2
\right ]  \lambda_0  
+ 6\,\lambda_2^2 \, ,\\
%%%
\label{l1}
\frac{d\lambda_1 } {dl} = & \left[ \epsilon
+ \left ( 2 \,N -1\right) \lambda _1+ \left ( 3 \,N-7 \right) \lambda _2
+ \frac{2 \left(\lambda _0 + \lambda _1 + 2 \, \lambda _2\right)}{N}
+ \left(\frac{14}{15}-\frac{2} {N}\right) u
\right ] \lambda_1
+ \frac{\lambda _0^2}{N} + \frac{34 \,\lambda _2^2}{5} \,,\\
%%%%%
\label{l2}
\frac{d\lambda_2 } {dl} = & 
 \left[ \epsilon +
\frac{ 13\,\lambda _0- 6 \,\lambda _1}  {5} 
+2 \,N\,\lambda _1
+ \frac{4 \left( 4 \,\lambda _1 - \lambda _0\right)}{N}
+\frac{ 
\left (60-16 \,N  \right ) u}  {15 \,N} \right ] \lambda _2
+  \left(N^2-7\, N-\frac{32}{N}  +26 \right) \lambda _2^2 \,,\\
%%%%%%%%%%%%%%%
\label{u}
\frac{du  } {dl} 
& =\left [\varepsilon 
+3 \times\frac{ \lambda_0 +  N\,   \lambda_1+ \frac{N \left( N-1\right ) \lambda_2} {2} } {2}
- \frac{8\, u}  {15}-2\,N_f \,u  \right  ]  u\,.
\end{align}

Let us examine these equations. Just as in Ref.~\cite{rahul-sid}, the flow $\frac{d\lambda_1 } {dl} $
for $\lambda_1$ continues to be strictly positive for a positive initial value of $\lambda_1$, and as a result, $\lambda_1$ grows under RG for ranges
encompassing small values of the coupling constants.
Next, note that $\lambda_2 = 0$ is a fixed point - if the action has time-reversal symmetry, the RG flow does not break it. Thus, the flow from Ref.~\cite{rahul-sid} is contained in the $\lambda_2=0$ subspace of the above equations. Moreover, if $\lambda_1$, $\lambda_2$, and $u$ start out from zero values,
these are driven to positive values by a positive $\lambda_1$, as long as the flowing coupling constants remain small enough to justify a perturbative treatment. We may, therefore, restrict our attention to regions of non-negative $\lambda_0$, $\lambda_1$, $\lambda_2$, and $u$.
%%%%%%%%%%%%%%%%%%%
Eventually, however, $\lambda_1$ must undergo a runaway flow to strong disorder, when the RG framework will break down. Consequently, there is no new fixed point at finite disorder emerging as a result of introducing $\lambda_2$, and the result, as in Ref.~\cite{rahul-sid}, is a runaway flow to strong disorder.

%Note also that   Eqs.~(\ref{l2})~and~(\ref{u}) are sign non-changing, \textit{i.e.} the $\beta$ function is proportional to the variable itself. As a result, if $\lambda_2$ and $u$ start out positive, they must remain non-negative. We will take non-negativity of $\lambda_2$ and $u$ for granted in the following discussion. Moreover, for $\lambda_2=0$, the flow reduces to that analyzed in Ref.~\cite{rahul-sid}. Since we are interested in searching for new physics, we assume {\it positivity} of $\lambda_2$.
%
%Now conditioned on positive $\lambda_2$, we can see by inspection that $\lambda_1$ cannot change sign as long as $\lambda_0$ is non-negative, whereas $\lambda_0$ cannot change sign as long as $\lambda_1$ is non-negative. Thus, if $\lambda_0$ and $\lambda_1$ start out positive, they can never become negative. Additionally, note that if either $\lambda_0$ or $\lambda_1$ (or both) start out zero, then they will be driven positive by $\lambda_2$. We thus conclude that we may restrict our attention to regions of positive $\lambda_0$, $\lambda_1$ and $\lambda_2$, and non-negative $u$. 
%
%Finally, note that for $N=5$, and conditioned on the positivity of the $\lambda_\alpha $'s and non-negativity of $u$, the $\beta$ function for $\lambda_1$ is {\it strictly positive}. This $\lambda_1$ must undergo a runaway flow to strong disorder. As such, there is no new fixed point at finite disorder emerging as a result of introducing $\lambda_2$, and the result, as in Ref.~\cite{rahul-sid}, is a runaway flow to strong disorder.

%%%%%%%%%%%%%%%%%%%%%%%%%%%%%%%%%%%%%%%%%%%%%%%%%%%%%%%%
\subsection{Strong-coupling trajectories}
\label{strong}

From Eq.~(\ref{l1}), we find that $\lambda_1$ has a strictly positive $\beta$ function, \textit{i.e.}, it is monotonically increasing under the RG flow. Therefore, we may view this as an RG time \cite{Vafek} such that we reparametrize the flows of $\lambda_{\alpha\neq 1}$ and $u$ in terms of $\lambda_1$. This gives us:
\begin{align}
\label{l02}
\frac{d\lambda_0} {d \lambda_1} = & \left[ \epsilon
-\frac{4 \left(15 \,N_f+4\right) u}{15}  + 2\, \lambda _0
+2 \left (N+1 \right) \lambda _1 +N \left (N-1 \right ) \lambda _2
\right ] \frac{\lambda_0}  {\frac{d \lambda_1  } {dl}} 
+ \frac{6\,\lambda_2^2} {\frac{d \lambda_1} {dl}} \, ,\\
%%%
\label{l22}
\frac{d\lambda_2 } {d \lambda_1} = & 
 \left[ \epsilon +
\frac{ 13\,\lambda _0- 6 \,\lambda _1}  {5} 
+2 \,N\,\lambda _1-\frac{4 \left(\lambda _0-4 \,\lambda _1\right)}{N}
+\frac{ 
\left (60-16 \,N  \right ) u}  {15 \,N} \right ] \frac{\lambda _2} {\frac{d \lambda_1} {dl}}
+  \left(N^2-7\, N-\frac{32}{N}  +26 \right) 
\frac{\lambda _2^2} {\frac{d \lambda_1} {dl}} \,,\\
%%%%%%%%%%%%%%%
\label{u2}
\frac{du } {d \lambda_1} 
& =\left [\varepsilon 
+3\times\frac{ \lambda_0
+  N\, \lambda_1+ \frac{N \left( N-1\right ) \lambda_2 } {2}   }{2}- \frac{8\, u}  {15}
-2 \,N_f \,u  \right ]  \frac{u}  {\frac{d \lambda_1} {dl}}\,,
\end{align}
where $ \frac{d \lambda_1} {dl}$ is obtained from Eq.~(\ref{l1}).

Observing that $\frac{\epsilon} {\lambda_1 }\rightarrow 0$ under the RG flow, in the trajectories towards strong coupling this `tree level' term is eventually unimportant, and we can simply look at the flow of ratios of couplings, viz. $\tilde \lambda_0 = \frac{\lambda_0} {\lambda_1} $, $ \tilde \lambda_2 = \frac{\lambda_2} {\lambda_1} $ and $ \tilde u = \frac{\lambda_2} {\lambda_1} $.  
The flows are then dictated by:
\begin{align}
\label{l0s1}
\frac{d \tilde \lambda_0} {d \ln \lambda_1} & \approx - \tilde \lambda_0 +
\frac{ \left[ 
-\frac{4 \left(15 \,N_f+4\right) \tilde u} {15}  + 2\,\tilde \lambda _0
+2 \left (N+1 \right)  +N \left (N-1 \right ) \tilde \lambda _2
\right ]  \tilde \lambda_0}  { den } 
+ \frac{6\,\tilde \lambda_2^2} { den } 
 \, , \\
%%%%%
\label{l2s1}    
\frac{d \tilde \lambda_2  } {d \ln \lambda_1} & \approx -\tilde \lambda_2 +
\frac{ \left[  
\frac{  13\, \tilde \lambda _0- 6 }  {5} 
+2 \,N -\frac{4 \left( \tilde \lambda _0-4  \right)}{N}
+\frac{ 
\left (60-16 \,N  \right ) \tilde u}  {15 \,N} \right ] \tilde \lambda _2} {den}
+ \frac{ \left(N^2-7\, N-\frac{32}{N}  + 26 \right) 
 \tilde \lambda _2^2} {den}\,,\\
%%%%%%%%%%%%%
\label{us1}
\frac{d   \tilde u  } {d \ln \lambda_1} 
& \approx -\tilde u +
\frac{ \left [ 
3\times\frac{ \tilde \lambda_0
+  N + \frac{N \left( N-1\right ) \tilde \lambda_2 } {2} }{2}
- \frac{8\,\tilde u}  {15}-2 \,N_f \,\tilde u  \right ]  
\tilde u}  {den}\,, 
\end{align}
where
\begin{align}
den & = 
\left[ -1-7 \,\tilde \lambda _2
+ \frac{2 \left( \tilde \lambda _0 + 1 + 2 \, \tilde \lambda _2\right)}{N}
+2 \,N \left( 1 + 3\,\tilde \lambda _2 \right)
+ \left(\frac{14}{15}-\frac{2} {N}\right) \tilde u
\right ] 
+ \frac{\tilde \lambda _0^2}{N} + \frac{34 \,\tilde \lambda _2^2}{5}
\,,
\end{align}
and we have set $\frac{\epsilon} {\lambda_1 } $ to zero.

For $\tilde \lambda_2 = 0$ these reduce to the flow equations from \cite{rahul-sid}. In that work two fixed points $(\tilde \lambda_0,  \tilde \lambda_2 = 0, \tilde u)$ were identified: the vector disorder only fixed point $(  0, \, 0, \, 0 )$, and the `scalar disorder dominated fixed point' $( 9.38516, \, 0, \,0)$. The former was unstable while the latter one was stable when the flow was restricted to the subspace of $\lambda_2=0$. 

What about non-zero $\lambda_2$? We have verified that there is no new fixed point at non-zero $\lambda_2$ for any value of $N_f $, i.e., the only fixed points are in the $\lambda_2 = 0$ subspace, given by
\begin{align}
\mathcal F_1=( 4+\sqrt{29}, \, 0, \,0), \,
\mathcal F_2=(   0, \, 0, \,0), 
\end{align}
corresponding to $(\tilde \lambda_0^*, \, \tilde \lambda_2^*,\, \tilde u^*)$.
The linearized flow equations in the vicinity of a fixed point are given by:
\begin{align}
\frac{d}{d\ln\lambda_1}
\left(\begin{array}{c} \delta  \tilde \lambda_0 \\ \delta \tilde \lambda_2 \\
\delta \tilde u   \end{array}\right) \Bigg\rvert _{(\tilde \lambda_0^*, \, \tilde \lambda_2^*,\, \tilde u^*)}
 \approx 
M
\left(\begin{array}{c} \delta  \tilde \lambda_0 \\ \delta \tilde \lambda_2 \\
\delta \tilde u   \end{array}\right),
\end{align}
where
\begin{align}
M =\begin{cases}
%%%%%%%%%
\left(
\begin{array}{ccc}
 - \frac{ 174 + 11 \,\sqrt{29} }  {355}  & 
 \frac{28 \left( 11 + 6 \,\sqrt{29} \right)}  {355}  & 
 \frac{-2 \left(  11 + 6 \,\sqrt{29} \right) \left (5 \,N_f + 2 \right )} {355}   \\
 0 & \frac{13} {47} & 0 \\
 0 & 0 & -\frac{19}{94} \\
\end{array}
\right)
 & \text{ for } \mathcal F_1 \,,\\
%%%%%%%%%%%%
\left(
\begin{array}{ccc}
 \frac{13}  {47} & 0 & 0 \\
 0 & \frac{13} {47} & 0 \\
 0 & 0 & -\frac{19}{94} \\
\end{array}
\right)
& \text{ for } \mathcal F_2 \,.
\end{cases}
\end{align}
The eigenvalues of $M$ for these two fixed points are given by:
\begin{align}
\left ( 
 -\frac{\left( 174 + 11 \sqrt{29}\right)}{355} , \,\frac{13} {47},
  \, -\frac{19}{94} \right )
\text{ and }  
 \left ( \frac{13}{47},\, \frac{13} {47},\,-\frac{19} {94}\right ),
\end{align}
respectively. The values show that $\mathcal{F}_1$ is stable. Another way to see this is to simply linearize the flow equation for $\tilde \lambda_2$ about the fixed points in the $\tilde \lambda_2 = 0$ subspace - it is straightforward to verify that $\tilde \lambda_2$ is an {\it irrelevant} perturbation, and so the flow to strong coupling is still controlled by the fixed trajectory $\mathcal{F_1}$, along which tensor disorder and Coulomb interactions vanish. 

We therefore conclude that time-reversal symmetry breaking `tensor' disorder is {\it irrelevant} in the sense that although it grows under renormalization group, its growth is asymptotically slower than the growth of time-reversal preserving scalar and vector disorder, such that the ratio of time-reversal breaking disorder strength to time-reversal preserving disorder strength flows to zero as the problem flows to strong disorder. This is the first main result of our work, and leads us to conjecture that the strong disorder physics should be dominated by time-reversal symmetry preserving disorder. Of course a rigorous treatment of the strong disorder physics is beyond the scope of a perturbative treatment in weak disorder as is employed in this work.

%%%%%%%%%%%%%%%%%%%%%%%%%%%%%%%%%%%%%%
\section{Unequal band masses}
\label{sec:ph-asym}

%---------------------------------------
%----------- FIGURE ---------------
%---------------------------------------
\begin{figure}
\includegraphics[width=0.45\linewidth]{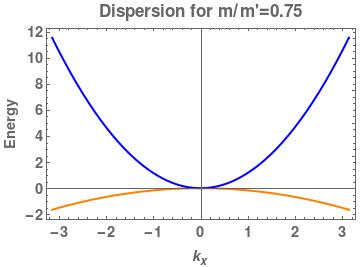}
\caption{Quadratic band touching with $m/m'=0.75$.
\label{fig:bands}}
\end{figure}

Thus far we have assumed that the conduction and valence bands have equal band mass. In this section we relax this assumption (while continuing to assume isotropicity). Specifically, we incorporate a scalar $\frac{k^2}{2\,m'}$ term in the bandstructure Hamilltonian, such that the bare Green's function becomes
\begin{equation}
\label{greennew}
G_0(\omega, \vec{k}) =  \frac{i \, \omega- \frac{k^2}{2\,m'}  + \dvec(\vec{k}) \cdot{\boldsymbol{\Gamma}}}
{-\left (i\, \omega- \frac{k^2}{2\, m'}  \right )^2 +|\dvec(\vec{k})|^2}\,.
\end{equation}
Now the two bands touching quadratically have different `curvatures' as shown in Fig.~\ref{fig:bands}. Note that we require $m' > m$ in order to be describing a quadratic band touching problem. For $m' < m$ both bands `curve' the same way, and for $m' = m$ one of the bands becomes perfectly flat, and neither of these cases is of interest to us here. Asymmetry of the band masses was shown to be irrelevant in the clean system \cite{Abrikosov}. However, we aim here to reassess its importance in the presence of disorder. 

\subsection{Renormalization of the band-mass asymmetry}
In the clean system, the self-energy coming from the Coulomb interaction takes the form:
\begin{align}
\Sigma(\omega, \vec{k}) & =-\frac{e^2}{c}   \int \frac{d\Omega \,d^d q}{(2\pi)^{d+1}} 
\frac{ \left [ i\, \omega +i\, \Omega - \frac{ (\vec{k} + \vec{q})^2}{2\,m'} + \dvec(\vec{k} + \vec{q}) \cdot{\boldsymbol{\Gamma}}\right]  V(q)}
{  - \left ( i\, \omega+i\, \Omega  - \frac{(\vec{k} + \vec{q})^2}{2\,m'} \right )^2 + |\dvec(\vec{k} + \vec{q})|^2} \nn
& = -\frac{e^2}{c}  \int \frac{d\Omega \,d^d q}{(2\pi)^{d+1}} 
\frac{\left [  i\, \Omega   + \dvec(\vec{k} + \vec{q}) \cdot{\boldsymbol{\Gamma}}  \right ] V(q)}
{  \Omega ^2 + |\dvec(\vec{k} + \vec{q})|^2} \,,
\end{align}
where by shifting the integration variable $\Omega$, we find that it gives the same correction of $\left [-\frac{m\, e^2
\, \dvec(\vec{k}) \cdot{\boldsymbol{\Gamma}}}{15\, \pi^2 \,c} \, \ln \left (  \frac {\Lambda_{UV} }  {\Lambda_{IR} }\right )   \right]$ as in the case with equal band masses. Hence, the one-loop renormalized Green's function becomes
\begin{align}
G^{-1} =&- i\, \omega +  \frac{k^2}{2\,m'} + \dvec(\vec{k}) \cdot{\boldsymbol{\Gamma}} \left(1 + \frac{m\, e^2}{15 \pi^2 \,c} l\right)
= - i\, \omega +  \frac{k^2}{2\,m'} + \frac{\tilde \dvec(\vec{k}) \cdot{\boldsymbol{\Gamma}} }
{ m \left(1 - \frac{8\, u}{15} l\right)}\,.
\end{align}
The requirement of Eq.~(\ref{rgm}) gives $z=2-\frac{8\,u}{15}$. Since $[m']= 2-z$, we find that $[m']$ has thus changed from $0$ at tree-level to $\frac{8\,u}{15}$ at one-loop level, i.e., the $\frac{k^2}{2\,m'}$ term becomes irrelevant, as anticipated in \cite{Abrikosov}. If we define the ratio $r_m =\frac{m}{m'}$ to parametrize the strength of band mass asymmetry ($r_m=0$ when electron and hole masses are equal),  then we conclude that that $[r_m]=-[m']=z-2<0$ in the clean system, i.e. $r_m$ flows to zero under RG.

Does disorder change this result? The self-energy contribution from the disorder terms is given by:
\begin{align}
\Sigma(\omega, \vec{k})  
 =& {2}\,W_0 \int \frac{d^d p}{(2\pi)^d} \, G(\omega, \vec{p}) 
   + 2\,W_1 \int \frac{d^d p}{(2\pi)^d} \,\Gamma_{a} G(\omega, \vec{p}) \Gamma_{a}
\nn & + 2\, W_2 \int  \frac{d^dp } {(2\, \pi)^d} \Gamma_{ab}  \, G (\omega,\mathbf k)\,  \Gamma_{ab} \nn
= &  \Big[ W_0 + N \, W_1+\frac{N\left(N-1\right) W_2}{2}\Big]
\,  \Big[    \frac{m^2 \, m'\, \Lambda_{\text{UV}}^2 } 
{4 \left (m^2 -m'^2  \right )\pi^2 } +
 \frac{ i\, \omega\, m^2\, m'^2 \left(m^2 + m'^2\right )  \ln \left (  \frac {\Lambda_{UV} }  {\Lambda_{IR} }\right ) }{\pi^2   \left(m^2 - m'^2\right )^2 }
\Big]\,.
\end{align}
The first term does not have any $\omega$ or $\vec k$ dependence and hence is just a chemical potential renormalization, which should be ignored assuming
we have the necessary correction to keep the system at the band-crossing point. Invariance of $[m]$ under the RG thus yields a dynamical exponent  
\begin{align}
z = &2 +  \frac{  m^2\, m'^2 \left(m^2 + m'^2\right ) \Big[ W_0 + N \, W_1+\frac{N\left(N-1\right) W_2}{2}\Big] }{ \pi^2  \left(m^2 - m'^2\right )^2 }
=  2+\frac{
m'^2 \left(m^2 + m'^2\right )\left[ \lambda_0 +  N \lambda_1 + \frac{N \left ( N-1\right) \lambda_2 } {2} \right]} 
{2   \left(m^2 - m'^2\right )^2}\,.
\end{align}
Now we have $z > 2$, such that the band mass asymmetry term becomes {\it relevant} under RG, in sharp contrast to the clean interacting case. 

What happens with both disorder and interactions? In this case we have 
\begin{align}
\label{dynexpcolnew}
z
= 2+ \frac{
 \left(1 + r_m^2\right )\left[ \lambda_0 +  N \lambda_1 + \frac{N \left ( N-1\right) \lambda_2 } {2} \right]} 
{2   \left(1 - r_m^2\right )^2}
- \frac{8\, u}{15} \,.
\end{align}
and $[r_m]=-[m']=z-2$, where recall $r_m = m/m'$ is zero for equal band masses. Now note that the interaction tries to make the band mass asymmetry irrelevant, but disorder makes it relevant - and recall also that disorder grows asymptotically more rapidly than interaction under the RG. We therefore conclude that as the problem flows to strong disorder, the strength of band mass asymmetry $r_m$ must grow. Eventually, there arises a scale where $r_m(l) = 1$. At this scale, one of the bands becomes {\it flat}, there arises a singularity in the density of states, and the whole RG scheme breaks down. We cannot push the RG beyond $r_m = 1$. Nevertheless, the prediction that band mass asymmetry should be {\it relevant} in the presence of disorder (whereas it was irrelevant in the clean system) is a non-trivial (and experimentally measurable) prediction of the RG, which should be apparent  in e.g. ARPES experiments. 

\subsection{Recomputing $\beta$ functions with unequal band masses}
In this section we recompute the $\beta$ functions with unequal electron and hole masses, still assuming $m' > m$ i.e. $r_m < 1$. This requires a re-evaluation of the integrals for all the constituent diagrams (but not a re-evaluation of combinatorial pre-factors or signs). 

\subsubsection{Clean system}
We begin with the clean system. For the diagram emerging from the contractions of the product of two Coulomb terms with the ZS topology, we obtain the contribution:
\begin{align}
&\Pi^{\text{ZS}}_{\text{cc}}(\bq) = - \frac{2 N_f}{q^4}\left(\frac{e^2}{2c}\right)^2
\Tr\Big[\int \frac{d^dk\, d\omega}{(2\pi)^{d+1}}\, 
 \frac{ \big \lbrace i\, \omega - \frac{\left(\mathbf k+\mathbf q \right)^2}{2\,m'}+ \dvec(\vec{k+q}) \cdot{\boldsymbol{\Gamma}} \big \rbrace \,
\big \lbrace    i\, \omega - \frac{k^2}{2\,m'}+ \dvec(\vec{k}) \cdot{\boldsymbol{\Gamma}}\big \rbrace   }
{  \Big \lbrace  - \left ( i\, \omega   - \frac{ (\vec{k+q})^2}{2\,m'} \right )^2 + |d (\vec{k+q})|^2  \Big  \rbrace   \,
\Big \lbrace - \left ( i\, \omega  - \frac{k^2}{2\,m'} \right )^2  + |d(\vec{k})|^2  \Big \rbrace} \Big]
\nonumber.
\end{align}
We choose $\vec{q}$ to lie along the $z$ axis, without any loss of generality.
Dropping the terms that will vanish upon performing the angular integrals, and performing the $\omega$ integral by the method of residues, we get
\begin{align}
\Pi^{\text{ZS}}_{\text{cc}}(\bq)=& -  \frac{ 8 N_f \,m'^2\,e^2}{c^2\,q^4} &  \int \frac{d^dk}{(2\pi)^d} \frac{ |\dvec(\vec{k+q})| + |\dvec(\vec{k})|}
{ 4  \left( |\dvec(\vec{k+q})| + |\dvec(\vec{k})| \right)^2 \,m'^2 +  \left( q^2+2\,k\,q \cos \theta \right)^2}
 \left( \frac{\dvec(\vec{k+q}) \cdot \dvec(\vec{k})}{|\dvec(\vec{k+q})||\dvec(\vec{k})|}-1\right).
\end{align}
Since the above integral manifestly vanishes for $\vec q=0$, we can obtain the divergent part from the leading order term in $q$ after Taylor expanding in small $q$, as follows:
\begin{eqnarray}
\Pi^{\text{ZS}}_{\text{cc}}(\bq) &=&    \frac{ 8 N_f  \,e^2}{c^2\,q^4}  \!\!\!\int \frac{d^dk}{(2\pi)^d}   \frac{3\,m\,q^2 }{4\, k^4} \sin^2\! \theta\,,
\label{picc}
\end{eqnarray}
This has the same form as in the case with equal band masses and gives the same correction of 
$\frac{e^2}{2\, c} \rightarrow \frac{e^2}{2\, c}   \left[ 1 - \frac{m \,e^2}{4\, \pi^2 \,c} N_f\,  l \right]$. As shown in Appendix~\ref{sec:CleanRGOnlyZS}, the remaining diagrams (VC, ZS$'$, BCS) do not have any divergent contribution and thus the band mass asymmetry does not affect the RG flows of the clean system.

%%%%%%%%%%%%%%%%%%%%%%%%%%%%%%%%%%%%%%%%%
\subsubsection{Disordered non-interacting system}

 A VC diagram with two scalar ($W_0$) lines can emerge in $8$ distinct ways; including factors from (\ref{disordercontraction}) we find a correction to the scalar vertex from
 \begin{align}
 \Gamma^{\text{VC}}_{00} =&  4 \,W_0^2
 \int\frac{d^d k}{(2\pi)^d} \frac{ \left [ \left (i\,\omega - \frac{k^2}{2\,m'^2} \right )+ \dvec_\vec{k}\cdot{\boldsymbol{\Gamma}}^{{j}} \right]
 }
{ \left [ -\left (i\,\omega - \frac{k^2}{2\,m'^2} \right )^2 + \frac{k^4}{4\,m^2} \right]^2 }
  \left [ \left (i\,\omega - \frac{k^2}{2\,m'^2} \right ) + \dvec_\vec{k} \cdot{\boldsymbol{\Gamma}}^{{j}} \right]\nn
&= 4 \,W_0^2   %
 \int\frac{d^d k}{(2\pi)^d} \frac{  \left (i\,\omega - \frac{k^2}{2\,m'^2} \right )^2 + \frac{k^4}{4\,m^2}  }
{\left [ -\left (i\,\omega - \frac{k^2}{2\,m'^2} \right )^2 + \frac{k^4}{4\,m^2} \right]^2 }
 =
\frac{2\, W_0^2 }
{  \pi^2}\, \frac{  m^2\, m'^2 \left(m^2 + m'^2\right )}{   \left(m^2 - m'^2\right )^2 }
\,  \ln \left ( \frac{\Lambda_{\text{UV}}}{\Lambda_{\text{IR}}}\right ).
 \end{align}

For the rest of the non-vanishing VC, BCS, and ZS$'$ diagrams we have to use the integral:
\begin{align}
 I  &= \int\frac{d^d k}{(2\pi)^d} \frac{  \frac{k^4}{4\,m^2}}
{ \left( -  \frac{k^4}{4\,m'^2}  +   \frac{k^4}{4\,m^2}  \right)^2}
=\frac{m^2\,m'^4 \,\ln \left ( \frac{\Lambda_{\text{UV}}}{\Lambda_{\text{IR}}}\right )}{2\, \pi^2\left (m^2-m'^2\right)^2 }\, ,
\label{int} 
 \end{align}
which appears as a prefactor for each. This can be implemented by replacing $m^2
 \rightarrow m^2\, \mu$, where $\mu\equiv\frac{ m'^4 }{ \left (m^2-m'^2\right)^2 } =\frac{ 1 }{ \left (1-r_m^2\right)^2 }$, in the answers obtained for the case with equal electron and hole masses.

%%%%%%%%
\subsubsection{Disordered interacting problem}
We now consider diagrams with mixed interaction and disorder lines. We start with the ZS diagram with one scalar disorder and one Coulomb line (all other ZS diagrams vanish upon taking the trace). We have
\be
\Pi^{\text{ZS}}_{c0} = -{2\, W_0}\times\frac{2\,c\,q^2}{e^2} \Pi^{\text{ZS}}_{cc} =  -\frac{N_f\,m\, e^2\,W_0\, \ln \left ( \frac{\Lambda_{\text{UV}}}{\Lambda_{\text{IR}}}\right )}{2\,\pi^2 \,c}
\,,
\ee
which is same as the case with equal band masses. 

We now consider vertex corrections. The Coulomb correction to a $W_0$ disorder vertex vanishes ($\Gamma^{\text{VC}}_{0c} = 0$) as before. The Coulomb correction to the $W_1$ vertex takes the form:
\begin{align}
\Gamma^{\text{VC}}_{1c} =& -  \frac{4\,W_1\,e^2}{2\,c}\Gamma^{{i}}_{a} 
 \int \frac{d \omega}{2\pi} \frac{d^d p}{(2\pi)^d}\, G(\omega, \vec{p})\,  \Gamma^{{j}}_{a} \,G(\omega, \vec{p}) \nonumber\\ 
=& \frac{2\,e^2\, W_1 \,  \Gamma^{{i}}_{a}\, \Gamma^{{j}}_a   }{c}
 \int \frac{d \omega}{2\pi} \frac{d^d p}{(2\pi)^d}\, \frac{-  \left (i\,\omega - \frac{ p^2}{2\,m'^2} \right )^2 + \frac{N-2}{N} \frac{p^4}
{4\, m^2} }
{p^{2}\left [  - \left (i\,\omega - \frac{p^2}{2\,m'^2} \right )^2+ \frac{p^4}{4\,m^2}  \right ]^2} \nn
=& \frac{2\,e^2\, W_1}{c} \Gamma^{{i}}_{a}\, \Gamma^{{j}}_a  
 \int \frac{d \omega}{2\pi} \frac{d^d p}{(2\pi)^d}\, \frac{ \omega^2 + \frac{N-2}{N} \frac{p^4}{4 \,m^2} }
{p^{2}\left [ \omega^2+ \frac{p^4}{4\,m^2}  \right ]^2} \,,
\end{align}
which is the same expression as in the equal mass case. Similarly, the Coulomb correction to the $W_2$ vertex also gives the same result as in Eq.~(\ref{gamma2c}).

Finally, we have:
\be
\Gamma^{\text{VC}}_{c0} = -\frac{e^2 \, \Gamma^{\text{VC}}_{00}}{2\,c \,q^2\, W_0}
 = - 
\frac{m^2\,m'^4\,e^2\, W_0 \, \ln \left ( \frac{\Lambda_{\text{UV}}}{\Lambda_{\text{IR}}}\right )}
{\left (m^2-m'^2\right)^2   c\, q^2\, \pi^2 } 
\,,
\ee
\be
\Gamma^{\text{VC}}_{c1} = -\frac{e^2 \, \Gamma^{\text{VC}}_{01}}{2\,c \,q^2 \,W_0} = -
\frac{m^2\,m'^4 \,N \,e^2\, W_1 \, \ln \left ( \frac{\Lambda_{\text{UV}}}{\Lambda_{\text{IR}}}\right )}
{\left (m^2-m'^2\right)^2 c\, q^2   \, \pi^2 } 
\,,
\ee
and
\be
\Gamma^{\text{VC}}_{c2} = -\frac{e^2 \,\Gamma^{\text{VC}}_{02}}{2\,c \,q^2 \,W_0} = -
\frac{ m^2\,m'^4\,  e^2 \,W_2 \,N \left ( N-1 \right ) \ln \left ( \frac{\Lambda_{\text{UV}}}{\Lambda_{\text{IR}}}\right )} 
{ 2\left (m^2-m'^2\right)^2 c\, q^2 \, \pi^2}
\,,
\ee

ZS$'$ and BCS diagrams with mixed disorder and interaction lines do not produce logarithmically divergent corrections, for the reasons identified in Ref.~\cite{rahul-sid}. 

\subsection{RG equations}

Firstly, we recall that in the presence of Coulomb interactions and disorder, the dynamical critical exponent is given by:
\begin{align}
\label{dynexpcolnew}
z
= 2+ \frac{
 \left(1 + r_m^2\right )\left[ \lambda_0 +  N \lambda_1 + \frac{N \left ( N-1\right) \lambda_2 } {2} \right]} 
{2   \left(1 - r_m^2\right )^2}
- \frac{8\, u}{15} \,.
\end{align}

Using Tables ~\ref{ZS}, \ref{VCnew}, and \ref{BCSnew}, we can write down the full set of the RG equations for the disorder couplings as well as $u$, when we include the $\frac{k^2}{2\,m'}$ term. We note that
%%%%%%%%%%%%%%%%%%%%%%%%
\begin{align}
\label{l0new}
\frac{d\lambda_0} {dl} &=
\left[ \varepsilon +2   \left( 1 +r_m^2\right) \mu\,\lambda _0  
+  \left\lbrace  2 +N \left( 2 +r_m^2\right) \right \rbrace \mu \, \lambda _1 
+\frac{ N\left (N-1 \right ) \left( 2 + r_m^2\right) \mu\,\lambda _2} {2}    
- \frac{ 4 \left( 4 + 15 N_f \right) u} {15}
\right]  \lambda_0 +  6 \, \mu\, \lambda _2^2 \, , \\
%%%
\label{l1new}
\frac{d\lambda_1 } {dl} &=
\left [ \varepsilon
+\frac{ \left( 2 + N\, r_m^2 \right ) \mu \,\lambda _0 } {N}
+   \left\lbrace  N \left( 2 +r_m^2 \right)+\frac{2}{N}
-1 \right \rbrace \mu\,\lambda _1 
+ \frac{  \left (N-1 \right ) \left(N^2 \,r_m^2+6 N-8\right) \mu\,\lambda _2 }  {2 N}
+ \left(\frac{14}{15}-\frac{2}{N}\right) u
\right ]  \lambda_1
\nn & \quad + \frac{\mu \, \lambda _0^2 } {N}
+ \frac{34 \,\mu \,\lambda_2^2 }{5} \,,   \\
%%%%%
\label{l2new}
\frac{d\lambda_2  } {dl} &=  \Big [ \varepsilon 
+\left(  r_m^2-\frac{4}{N}+ \frac{13}{5} \right) \mu\,\lambda_0
+ \left \lbrace
N \left( 2 +r_m^2 \right)+\frac{16}{N}-\frac{6} {5}
\right \rbrace \mu\,\lambda_1
%%%
+ \frac{ 52 +N \left (N-1 \right ) r_m^2
+2 \,N\left (N-7 \right ) 
-\frac{64}{N}} {2} \,\mu\,\lambda_2  \nn & \qquad
+\left(\frac{4}{N}-\frac{16}{15}\right) u
\Big ]  \lambda_2  \,,    \\
%%%%%%%%%%%%%
\label{urgnew}
\frac{du  } {dl} 
& =
\left [ \varepsilon +\left ( 3  +r_m^2 \right )\times
\frac{ \lambda_0+   N\,   \lambda_1+ \frac{N \left( N-1\right ) \lambda_2  } {2}   }
{ 2 } \,\mu
- \frac{8\, u}{15}-2N_f \,u  \right  ]  u\,,  \\
%%%%%%%%%%%%%
\label{rmnew}
\frac{d r_m  } {dl} 
& =  \left [ 
\frac{   \left( 1 + r_m^2\right ) 
\left \lbrace \lambda_0 +  N \lambda_1 
+ \frac{N \left ( N-1\right) \lambda_2 } {2} \right \rbrace \mu} {2 } 
- \frac{8\, u}{15}  \right] r_m\,.
\end{align}
%%%%%%%%%%%%
These equations reduce to the equations for equal electron and hole mass when we take $r_m \rightarrow 0$ and $\mu \rightarrow 1$.

\begin{table}
\begin{tabular}{|c|c|c|c||c|}
\hline
Coupling  & $ \lambda_0$ & $ \lambda_1 $ & $\lambda_2$ & $u$ \tabularnewline
\hline
$ \lambda_0$ & $\delta \lambda_0 =\left (1+r_m^2\right )  \mu\, \lambda_0^2  \, l $ & $\delta \lambda_1 = -   \frac{N-2}{N }\,  \mu\, \lambda_0 \, \lambda_1\, l $  &   
$\delta \lambda_0 =\frac{ N \left ( N-1 \right ) }{2}     \mu\,\lambda_0\, \lambda_2 \, l$
 & 0 \tabularnewline
\hline
$ \lambda_1$ &$\delta \lambda_0 =  N\,   \mu\,\lambda_0\, \lambda_1\,l$&$\delta \lambda_1 =  \frac{(N-2)^2}{ N}\,  \mu\, \lambda_1^2\, l$& 
$ \delta \lambda_1   = - \frac{ \left ( N-1 \right )  \left ( N-2 \right )  \left (N-4 \right ) }{ 2 N}\,  \mu\, \lambda_1\, \lambda_2 \, l  $
&$\delta  \lambda_1 =    \frac{2 \left (N-1 \right )}{N } \, \lambda_1\, u  \,l$ \tabularnewline
\hline
$ \lambda_2$ &  $\delta \lambda_2 = \frac{  N-4 }{N}\,  \mu\, \lambda_0 \,\lambda_2 \, l $
& $\delta \lambda_2 = \frac{  \left(N-4 \right )^2}{N}\,   \mu\,\lambda_1 \,\lambda_2 \, l $&
 $ \delta \lambda_2 = \frac{ \left (  N-4 \right ) \left (N^2-9N + 16 \right ) }{2 N}\,  \mu\, \lambda_2^2 \, l $
& $d\lambda_2=\frac{4 } {N}  \,\lambda_2\, u  \, l $ \tabularnewline
\hhline{|=|=|=|=||=|}
$u$ & $\delta u =  \mu\, \lambda_0\, u \, l $ &$ \delta u =  N \,  \mu\,\lambda_1 \,u \, l$
& 
$ \delta u= \frac{  N\left ( N-1 \right )} { 2   }\,    \mu\,  \lambda_2 \,u\, l$
 &0\tabularnewline
\hline
\end{tabular}
\caption{Contributions to the $\beta$-functions from the VC diagrams with the $\frac{k^2}{2\,m'}$ term. Here, $\lambda_\alpha = \frac{2\,m^2\,W_\alpha}{\pi^2}$, $u = \frac{m\,e^2}{8 \,\pi^2\, c}$, $\mu=  \frac{ 1 }{ \left (1-r_m^2\right)^2}$ and $l$ is the RG flow parameter. \label{VCnew}} 
\end{table}

%%%%%%%%%%%%%%%%%%%%%%%%%%%%%%%
 \begin{table}
\begin{tabular}{|c|c|c|c||c|}
\hline
Coupling  & $ \lambda_0$ & $ \lambda_1 $ & $\lambda_2 $ & $u$ \tabularnewline
\hline
$ \lambda_0$ &
$\delta \lambda_1 = \frac{1}{N}\, \mu  \,\lambda_0^2\, l$& 
$\delta \lambda_0 = 2\, \mu  \, \lambda_0\, \lambda_1\,l$
&
$\delta \lambda_2 =\frac{3} {5}\,\mu\,\lambda_0\, \lambda_2  \, l$
& $ 0$
 \tabularnewline
\hline
$\lambda_1$ &
included in $(\lambda_0, \,\lambda_1)$ cell &
$ \delta \lambda_1 =   \frac{3N-2}{N}\, \mu  \,\lambda_1^2\, l$ & 
$\delta \lambda_2 =\frac{34} { 5 }\,\mu\, \lambda_1\, \lambda_2  \, l$
&$0$ \tabularnewline
\hline
%%%
$\lambda_2 $ &included in $(\lambda_0, \,\lambda_2)$ cell
& included in $(\lambda_1, \,\lambda_2)$ cell &
%%%%%%%%%%%%%%%%%
\makecell{$\delta \lambda_0 = 6 \, \mu\,\lambda_2^2 \, l $,\\  
$\delta \lambda_1 = { \frac{ 34 } {5} } \, \mu\,\lambda_2^2 \, l $}
%%%%%%%%%%%%%
& $0$ \tabularnewline
\hhline{|=|=|=|=||=|}
$u$ &  $0$ & $0$ &     $0$ &  $0$\tabularnewline
\hline
\end{tabular}
\caption{Sum of contributions to the $\beta$-functions from the BCS and ZS$'$ diagrams with the $\frac{k^2}{2\,m'}$ term, using the same conventions as Table~\ref{VCnew}. \label{BCSnew}}
\end{table}
%%%%%%%%%%%%%%%%%%%%%%%%%%%%%%%%%

Now let us discuss the fixed point structure of the above equations. Firstly, note that the $\beta$ functions for $\lambda_2$ and $u$ are proportional to the variables themselves, and so these variables cannot change sign, nor can they be generated `from nothing' under RG. If they start out positive, they must remain positive forever. Next, note that (given the non-negativity of $\lambda_2$), it follows from arguments analogous to those advanced in the case with equal masses that $\lambda_0$ and $\lambda_1$ cannot become negative, if they start out with non-negative initial values. We thus conclude that all four couplings $\lambda_{1,2,3}$ and $u$ must be non-negative. 

Now note that for non-negative couplings, with $N=5$ and $0\le r_m<1$, the beta function for $\lambda_1$ is strictly positive, so that $\lambda_1$ must grow without limit i.e. there is not any finite disorder fixed point of the perturbative RG, even after we allow for unequal band masses. Thus the band mass asymmetric problem also flows to strong disorder. 

Finally, a consideration of the equation for $r_m$ with growing $\lambda_i, u$ leads to the conclusion that there is no fixed point at $0 < r_m < 1$. The only fixed point for this equation is at $r_m=0$, and this fixed point is {\it unstable} in the presence of disorder (as has been discussed). Thus, $r_m$ increases without limit under perturbative RG (although the RG scheme itself starts to break down when $r_m \rightarrow 1$, at which point one of the two bands becomes flat, and the co-efficient $\mu$ becomes singular). 

%%%%%%%%%%%%%%%%%
\subsection{Strong-coupling trajectories}
\label{strong2}

From Eq.~(\ref{l1new}), we find that $\lambda_1$ has a strictly positive $\beta$ function, \textit{i.e.}, it is monotonically increasing under the RG flow. Therefore, we may view this as an RG time such that we reparametrize the flows of $\lambda_{\alpha\neq 1}$, $u$ and $r_m$ in terms of $\lambda_1$. This gives us:
%%%%%%%%%%%%%%%%%%%%%%%%
\begin{align}
\frac{d\lambda_0} {d  \lambda_1} &=
\frac{
\left[ \varepsilon +2   \left( 1 +r_m^2\right) \mu\,\lambda _0  
+  \left\lbrace  2 +N \left( 2 +r_m^2\right) \right \rbrace \mu \, \lambda _1 
+\frac{ N\left (N-1 \right ) \left( 2 + r_m^2\right) \mu\,\lambda _2} {2}    
- \frac{ 4 \left( 4 + 15 \,N_f \right) u} {15}
\right]  \lambda_0 +  6 \, \mu\, \lambda _2^2
}
{\frac{d \lambda_1  } {d  l}}
 \, ,\label{l0new1} \\
%%%%%
\frac{d\lambda_2  } {d   \lambda_1} &=
\frac{ \left[ \varepsilon 
+\left(  r_m^2-\frac{4}{N} + \frac{13} {5} \right) \mu\,\lambda_0
+ \left \lbrace
N \left( 2 +r_m^2 \right)+\frac{16}{N}-\frac{6}{5}
\right \rbrace \mu\,\lambda_1
%%%
+ \frac{ 52 +N \left (N-1 \right ) r_m^2
+2 \,N\left (N-7 \right ) 
-\frac{64}{N}} {2} \,\mu\,\lambda_2 
+\left(\frac{4}{N}-\frac{16}{15}\right) u
\right ]  \lambda_2 }
{\frac{d \lambda_1  } {d  l}}\,,
\label{l2new1}    \\
%%%%%%%%%%%%%
\frac{du  } {d  \lambda_1} 
& = \frac{
\left [\varepsilon +  \left ( 3  +r_m^2 \right ) \times
\frac{ \lambda_0+   N\,   \lambda_1+ \frac{N \left( N-1\right ) \lambda_2  } {2}   }
{ 2 } \,\mu
- \frac{8\, u}{15}-2N_f \,u 
 \right ]  u}  {\frac{d \lambda_1  } {d  l}}\,, \label{urgnew1}  \\
%%%%%%%%%%%%%
\frac{d r_m  } {d   \lambda_1} 
& =  \frac{ \left [ 
\frac{   \left( 1 + r_m^2\right ) 
\left \lbrace \lambda_0 +  N \,\lambda_1 
+ \frac{N \left ( N-1\right) \lambda_2 } {2} \right \rbrace \mu  } 
{2 }-  \frac{8\, u}  {15}  \right] r_m} 
{ \frac{d \lambda_1  } {d  l}  }\,,
\label{rmnew1}
\end{align}
where $ \frac{d \lambda_1  } {d  l}$ is obtained from Eq.~(\ref{l1new}).

Observing that $\frac{\epsilon} {\lambda_1 }\rightarrow 0$ under the RG flow, in the trajectories towards strong coupling this `tree level' term is eventually unimportant, and we can simply look at the flow of ratios of couplings, viz. $\tilde \lambda_0 = \frac{\lambda_0} {\lambda_1} $, $ \tilde \lambda_2 = \frac{\lambda_2} {\lambda_1} $ and $ \tilde u = \frac{\lambda_2} {\lambda_1} $.  
The flows are then dictated by:
%%%%%%%%%%%%%%%%%%%%%%%%
\begin{align}
\frac{d\lambda_0} {d \ln  \lambda_1}
 & \approx  -\tilde\lambda_0 
+ \frac{\left[  2   \left( 1 +r_m^2\right) \mu\,\tilde \lambda _0  
+  \left\lbrace  2 +N \left( 2 +r_m^2\right) \right \rbrace \mu  
+\frac{ N\left (N-1 \right ) \left( 2 + r_m^2\right) \mu\,\tilde \lambda _2} {2}    
- \frac{ 4 \left( 4 + 15 N_f \right) \tilde u} {15}
\right]  \tilde \lambda_0}   {den'}
 +  \frac{ 6 \, \mu\,\tilde \lambda _2^2 }  {den'}
 \, ,\label{l0new2} \\
%%%%%
\frac{d\lambda_2 } {d\ln \lambda_1} &  \approx  -\tilde \lambda_2
+ \frac{
\Big [ \left(  r_m^2-\frac{4}{N}+\frac{23}{10} \right) \mu\,\tilde \lambda_0
+ \left \lbrace
N \left( 2 +r_m^2 \right)+\frac{16}{N}-\frac{23}{5}
\right \rbrace \mu
%%%
+ \frac{ 52 +N \left (N-1 \right ) r_m^2
+2 \,N\left (N-7 \right ) 
-\frac{64}{N}} {2} \,\mu\,\tilde \lambda_2  
+\left(\frac{4}{N}-\frac{16}{15}\right) \tilde u
\Big ] \tilde  \lambda_2
}
{den'}
\,,
\label{l2new1}    \\
%%%%%%%%%%%%%
\frac{du  } {d \ln  \lambda_1} 
&  \approx  -\tilde u+
\frac{ \left[ 
\left (3+r_m^2 \right ) \times
\frac{ \lambda_0+   N+ \frac{N \left( N-1\right )  \tilde \lambda_2  } {2}   }
{ 2 } \,\mu
- \frac{8\,  \tilde u}{15}-2N_f \, \tilde u 
 \right ]  \,\tilde u}    {den' }\,, \label{urgnew2}  \\
%%%%%%%%%%%%%
\frac{d r_m  } {d   \ln \lambda_1} 
&  \approx  \frac{  \left [ 
\frac{   \left( 1 + r_m^2\right ) \big \lbrace \tilde \lambda_0 +  N  + \frac{N \left ( N-1\right) \tilde  \lambda_2 } {2} \big\rbrace \mu }  {2 }
-  \frac{8\, \tilde u}{15}  \right] r_m
}
{den'}
\,,
\label{rmnew2}
\end{align}
where
\begin{align}
den' &=
\frac{ \left( 2 + N\, r_m^2 \right ) \mu \,\tilde \lambda_0 } {N}
+   \left\lbrace  N \left( 2 +r_m^2 \right)+\frac{2}{N}
-1 \right \rbrace \mu  
+ \frac{  \left (N-1 \right ) \left(N^2 \,r_m^2+6 N-8\right) \mu\,\tilde \lambda _2 }  {2 N}
+ \left(\frac{14}{15}-\frac{2}{N}\right) \tilde u
 \nn &  \qquad + \frac{\mu\, \tilde \lambda_0^2} {N}
+ \frac{34 \,\mu \,\tilde \lambda_2^2 }{5}\,,
\end{align}
and we have set $\frac{\epsilon} {\lambda_1 } $ to zero.

For any $N_f  $, we obtain the following non-negative fixed points:
\begin{align}
\mathcal F_1=( 4+\sqrt{29}, \, 0, \,0,0), \,
\mathcal F_2=(   0, \, 0, \,, 0), 
\end{align}
corresponding to $(\tilde \lambda_0^*, \, \tilde \lambda_2^*,\, \tilde u^*, \, r_m^*)$. These are the same fixed points that were obtained in the case with equal band masses. It follows from our earlier analysis that $\mathcal F_1$ is stable in the  $(\tilde \lambda_0^*, \, \tilde \lambda_2^*,\, \tilde u^*)$ subspace, whereas $\mathcal F_2$ is unstable. However, both fixed points are unstable in the $r_m$ direction.
A breakdown of the RG procedure when $r_m \rightarrow 1$.

%---------------------------------------
%----------- FIGURE ---------------
%---------------------------------------
\begin{figure}
\includegraphics[width=0.45\linewidth]{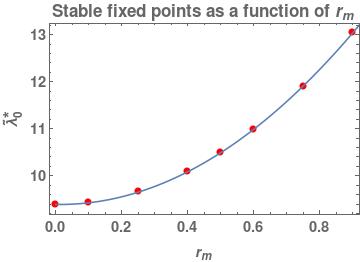}
\caption{Behaviour of the scalar-vector disorder ratio as a function of $r_m$, at the stable fixed point of the ratio space.
\label{sfp}}
\end{figure}

We have also calculated the fixed points in the ratio space as functions of $r_m$ (treating $r_m$ as a fixed rather than flowing parameter) in order to illuminate how the asymmetry in band masses affects the RG flow. The nature of the fixed points are still of the form $(   {\tilde{ \lambda}_0}^*,0,0),\,(0,0,0)$, with a positive ${\tilde{ \lambda}_0}^*$. The first fixed point is stable, while the second one is unstable. The behaviour of ${\tilde{ \lambda}_0}^*$ as a function of $r_m$ is shown in Fig.~\ref{sfp}. This gives us a sense of how the flow in the coupling space changes as band mass asymmetry becomes strong. Importantly, while the ratio of scalar to vector disorder changes (so that scalar disorder becomes more important as the band mass asymmetry becomes strong), tensor disorder and the long range interaction continue to grow asymptotically more slowly than scalar and vector disorder, and may continue to be ignored in a first approximation. 

%%%%%%%%%%%%%%%%%%%%%%%%%%%%%%%%%%%%%%%%%%%

\section{Analysis and discussion \label{sec:analysis}}
We have analyzed the interplay of short-range disorder and Coulomb interactions about quadratic band crossings in three dimensions, using a perturbative renormalization group procedure. Unlike earlier work \cite{rahul-sid, LaiRoyGoswami}, we have not restricted ourselves to time-reversal symmetry preserving disorder, nor have we assumed that the conduction and valence bands have equal mass. (Time reversal symmetry breaking disorder may come physically from e.g. magnetic impurities). We have shown that the full problem, including all types of disorder as well as unequal band masses, does not admit any non-trivial stable fixed points at weak coupling, and exhibits a runaway flow to strong disorder. Along the flow to strong disorder, time-reversal-symmetry-preserving disorder grows asymptotically faster than time-reversal-symmetry-breaking disorder and the Coulomb interaction. Thus, we conjecture that at a first pass, both time-reversal-symmetry-breaking disorder and Coulomb interactions may be neglected in describing the strong-coupling phase, and only time-reversal-preserving disorder needs to be taken into account. In this respect, the general problem that we study herein `flows' into the simpler problem tackled in Ref.~\cite{rahul-sid}, and the discussion therein regarding the strong-coupling phase may be carried over {\it mutatis mutandis}, and two phases (a diffusive metal and a localized phase) may be predicted. The critical point between the localized and diffusive phases would serve as an interesting test bed for a {\it many body localization} transition, insofar as the bare Hamiltonian is interacting, even if disorder is more relevant than the interaction.  Of course, arguments about the strong-coupling regime based on extrapolation from weak coupling must be treated with caution, and a careful discussion of the strong-coupling physics would require construction of the appropriate sigma model. Furthermore, even if time-reversal-symmetry-breaking disorder is asymptotically weaker than time-reversal-symmetry-preserving disorder, it may still have important effects by changing the symmetry class of the problem. Nevertheless, our results do suggest that in describing the strong-coupling phase, it should be sufficient to {\it start} with an analysis of the effects of strong time-reversal-symmetry-preserving disorder, and then to incorporate time-reversal-symmetry-breaking disorder and the Coulomb interaction as perturbations. Construction of such a description of the strong-coupling phase would be an interesting challenge for future work. 

Remarkably, our analysis also reveals that whereas band mass asymmetry is {\it relevant} in the presence of disorder. This is in sharp contrast to the situation that appears in clean systems \cite{Abrikosov}, where asymmetry of the band masses is {\it irrelevant}. This distinction between clean and dirty systems constitutes the most important prediction of our work  - we predict that whereas in clean systems the conduction and valence bands should have the same mass in the scaling limit, in dirty systems they should have very different masses (see e.g. Fig.~\ref{fig:bands}), and indeed in the low energy limit one of the two bands should become asymptotically flat. This is a non-trivial prediction of our analysis, which could be directly probed in, e.g., ARPES experiments, and would provide a {\it direct} experimental diagnostic of whether disorder physics dominates a particular sample, or whether the sample may be treated as `effectively clean.'  Experimental investigations of such systems are just starting \cite{Armitage}. We hope that our work will prove useful in guiding experiments, as they seek to explore this novel regime. 

{\bf Acknowledgements} R.M.N. would like to thank S.A. Parameswaran for a previous collaboration on a related problem, and Y.Z. Chou for useful discussions. We also acknowledge an illuminating conversation with Philipp Dumitrescu. R.M.N.'s research was sponsored by the U.S. Army Research Office and was accomplished under Grant Number W911NF-17-1-0482. The views and conclusions contained in this document are those of the authors and should not be interpreted as representing the official policies, either expressed or implied, of the Army Research Office or the U.S. Government. The U.S. Government is authorized to reproduce and distribute reprints for Government purposes notwithstanding any copyright notation herein.

%%%%%%%%%%%%%%%%%%%%%%%%%%%%%%5
\appendix
\section{Clifford algebra and various identities}
\label{app1}

 In this appendix, we list various identities which follow from the Clifford algebra. First, for $N$ gamma matrices ${\Gamma_a}$ $(a=1,2,\ldots, N)$, we have
\begin{align}
\sum \limits_{a} \Gamma_a\,\Gamma_a =N\,.
\label{rel0}
\end{align}
Other relations that have been used in various computations in the main text are:
\begin{align}
\label{rel0}
& \sum \limits_{a<b } \Gamma_{ab}\, \Gamma_{ab} =\frac{N \left(N-1\right )}{2} \,,
\\
\label{rel1}
& \Gamma_{cd}\, \Gamma_a=2\, i \left ( \delta_{ac}   \,\Gamma_{d}- \delta_{ad} \, \Gamma_c \right )
+\Gamma_a \, \Gamma_{cd}  \,, \\
%%%%%%%%%
\label{rel2}
 & \Gamma_{cd}\,  \Gamma_{a}\,  \Gamma_{cd}  =\frac{\left (N-1\right ) \left (N-4\right )}{2}  
 \Gamma_{a}\,, \\
 \label{rel3}
&\Gamma_{f}\,\Gamma_{ab}\, \Gamma_{f}=\left (N-4\right ) \Gamma_{ab}\,,
\\ \label{rel4}
& \Gamma_{cd}\,\Gamma_{ab}\,\Gamma_{cd} = \frac{ N^2-9\, N +16} {2}
\, \Gamma_{ab}\,.
\end{align}

Since $N=5$ for the current problem, we can use the relations:
\begin{align}
& \Gamma_{a b}\, \Gamma_f =   
 i \left (  \delta_{a f} \, \Gamma_b - \delta_{b f} \,\Gamma_a \right ) - 
  \frac{ \epsilon_{abfcd}\, \Gamma_{cd} }{2} \,,
\quad
\Gamma_f \,\Gamma_{ab} = 
i \left (  \delta_{b f} \,\Gamma_a -\delta_{a f} \, \Gamma_b \right ) 
- \frac{ \epsilon_{abfcd}\, \Gamma_{cd} }{2}  \,,
%%%%%%%%%%%%%%%%%%%%
\\ &  \epsilon_{abcde}\,\epsilon_{abklm}
= 6 \left(  \delta_{dl}\,\delta_{em} -\delta_{dm} \,\delta_{el} \right),
%%%%%%%%%%%%%%%%%%%%%%%%%%
\\&  \epsilon_{abcde}\,\epsilon_{abclm} =
2 \left(    \delta _{cl}\, \delta _{ek}\, \delta _{dm}
+\delta _{ck}\, \delta _{dl} \, \delta _{em}
-\delta _{ck}\, \delta _{dm}\, \delta _{el}
-\delta _{cl}\, \delta _{dk}\, \delta _{em}
+\delta _{cm}\, \delta _{dk}\, \delta _{el}
-\delta _{cm}\, \delta _{dl}\, \delta _{\text{ek}} \right).
\label{rel41}
\end{align}
Using these, we get:
\begin{align}
\label{rel51}
\sum \limits_{a <  b} \Gamma^i_{ab}\,  \Gamma^i_{e } 
\left (  \Gamma^j_{ab} \,  \Gamma^j_{e}  +  \Gamma^j_{e}  \,  \Gamma^j_{ab}\right )
& = 3 \sum \limits_{a< b} \Gamma_{ab}^i\, \Gamma_{ab}^j \,,
%%%%%%%%%%%%%%%%%%%%%%%%%
 \\ \label{rel52}\sum \limits_{a <  b,\,c, \, f} \Gamma^i_{ab}\,  \Gamma^i_{f} \,  \Gamma^i_{ c } 
\left ( \Gamma^j_{ab}  \,  \Gamma^j_{f} \,  \Gamma^j_{c }
+ \Gamma^j_{c } \,  \Gamma^j_{f} \,  \Gamma^j_{ab}  \right )
&= 34 \sum \limits_{a< b} \Gamma_{ab}^i\, \Gamma_{ab}^j \,,
%%%%%%%%%%%%%%%%%%%%
\\ \label{rel53}
\sum \limits_{a <  b,\,c<d, \, e} \Gamma^i_{ab}\,  \Gamma^i_{e} \,  \Gamma^i_{ cd } 
\left (  \Gamma^j_{ab } \,  \Gamma^j_{e} \,  \Gamma^j_{cd}  
+  \Gamma^j_{cd}  \,  \Gamma^j_{e} \,  \Gamma^j_{ab}\right )
& =  60 + 68 \sum \limits_{a} \Gamma_{a}^i\, \Gamma_{a}^j\,.
\end{align}

%%%%%%%%%%%%%%%%%%%%%%%

\section
{Vanishing divergent contributions from the VC, BCS, ZS$'$ diagrams in the clean system with $\frac{k^2}{2\,m'}$ term}
\label{sec:CleanRGOnlyZS}

For the clean system with equal electron and hole masses, it was shown in Ref.~\cite{rahul-sid} that the VC, BCS, and ZS$'$ diagrams give no divergent contribution. Here we verify that the above statement holds even in the presence of the $\frac{k^2}{2\,m'}$ term.

The vertex correction with two Coulomb lines has a relative minus sign compared to ZS and takes the form:
\begin{align}
\Gamma^{\text{VC}}_{00} &\propto - \frac{  e^4}{c^2\,q^{2}}\int d\omega \,d^d k 
\, \frac{G(\omega, \vec{p}) G(\omega, \vec{p}+\vec{q})}{k^{2}} \nn
&
\propto - \frac{  e^4}{c^2\,q^{2}}\int d\omega \,d^dk \,
 \frac{ \big \lbrace i\, \omega - \frac{\left(\mathbf k+\mathbf q \right)^2}{2\,m'}+ \dvec(\vec{k+q}) \cdot{\boldsymbol{\Gamma}} \big \rbrace \,
\big \lbrace    i\, \omega - \frac{k^2}{2\,m'}+ \dvec(\vec{k}) \cdot{\boldsymbol{\Gamma}}\big \rbrace   }
{ k^2 \, \Big \lbrace  - \left ( i\, \omega   - \frac{ (\vec{k+q})^2}{2\,m'} \right )^2 + |d (\vec{k+q})|^2  \Big  \rbrace   \,
\Big \lbrace - \left ( i\, \omega  - \frac{k^2}{2\,m'} \right )^2  + |d(\vec{k})|^2  \Big \rbrace}\,.
\end{align} 
We choose $\vec{q}$ to lie along the $z$ axis, without any loss of generality.
Dropping the terms that will vanish upon performing the angular integrals, and performing the $\omega$ integral by the method of residues, we get
\begin{align}
\Gamma^{\text{VC}}_{00} &\propto
- \frac{  e^4}{c^2\,q^{2}}  \int \frac{d^dk}{k^2} \frac{ |\dvec(\vec{k+q})| + |\dvec(\vec{k})|}
{ 4  \left( |\dvec(\vec{k+q})| + |\dvec(\vec{k})| \right)^2 \,m'^2 +  \left( q^2+2\,k\,q \cos \theta \right)^2}
 \left( \frac{\dvec(\vec{k+q}) \cdot \dvec(\vec{k})}{|\dvec(\vec{k+q})||\dvec(\vec{k})|}-1\right).
\end{align}
Since the above integral manifestly vanishes for $\vec q=0$, we can obtain the possible divergent part from the leading order term in $q$ after Taylor expanding in small $q$, as follows:
\begin{align}
\Gamma^{\text{VC}}_{00} &\propto - \frac{  e^4}{c^2 } 
\int {d^dk} \, \frac{3\,m\, \sin^2\! \theta }{4\, k^ 6 } \,,
\label{picc}
\end{align}
which diverges as $1/q^2$ i.e produces a correction to the Coulomb line, but only a constant correction (since there is no log divergence). Thus, this diagram does not contribute the the $\beta$ functions of the clean system. 

The ZS$'$ and BCS diagrams correspond to the ladder and twisted ladder diagram topologies (``Cooperon'' and ``diffuson'').
Denoting the incoming momenta by $\vec{k_1}$ and $\vec{k_2}$, and the outgoing momenta by $\vec{k_1} + \vec{q}$ and $\vec{k_2} - \vec{q}$, we note that for extracting the possible divergent part, although the external momenta $\vec k_1$ and $\vec k_2$ can be set to zero, the momentum transfer $\vec{q}$ cannot be, as it is needed to split a high order pole in the integrand coming from the doubled Coulomb line \cite{NandkishoreLevitov}. However, we can choose $\vec{q}$ to lie along the $z$ axis, without any loss of generality. The sum of these two diagrams with two Coulomb lines gives (note the overall minus sign with respect to ZS):
\begin{align}
 \Pi^{\text{ZS}'}_{00} + \Pi^{\text{BCS}}_{00} 
 &\propto  - e^4\int \frac{d \omega\, d^d k}{k^{2}\, |\vec{k}-\vec{q}|^{2}} G(\omega, \vec{k}) \left[G(\omega, \vec{k} - \vec{q}) +   G(-\omega, - \vec{k})\right] \\ 
&  \propto  - e^4\int \frac{d \omega\, d^d k}{k^{2}\, |\vec{k}-\vec{q}|^{2}}
\Big [
\frac{ \big \lbrace i\, \omega - \frac{\left(\mathbf k+\mathbf q \right)^2}{2\,m'}+ \dvec(\vec{k+q}) \cdot{\boldsymbol{\Gamma}} \big \rbrace \,
\big \lbrace    i\, \omega - \frac{k^2}{2\,m'}+ \dvec(\vec{k}) \cdot{\boldsymbol{\Gamma}}\big \rbrace   }
{   \Big \lbrace  - \left ( i\, \omega   - \frac{ (\vec{k+q})^2}{2\,m'} \right )^2 + |d (\vec{k+q})|^2  \Big  \rbrace   \,
\Big \lbrace - \left ( i\, \omega  - \frac{k^2}{2\,m'} \right )^2  + |d(\vec{k})|^2  \Big \rbrace}
\nn & \hspace{ 4.3 cm}
-  \frac{ \big \lbrace i\, \omega - \frac{   k   ^2}{2\,m'}+ \dvec(\vec{k }) \cdot{\boldsymbol{\Gamma}} \big \rbrace \,
\big \lbrace    i\, \omega + \frac{k^2}{2\,m'}+ \dvec(\vec{k}) \cdot{\boldsymbol{\Gamma}}\big \rbrace   }
{   \Big \lbrace  - \left ( i\, \omega   - \frac{ k^2}{2\,m'} \right )^2 + |d (\vec{k })|^2  \Big  \rbrace   \,
\Big \lbrace - \left ( i\, \omega + \frac{k^2}{2\,m'} \right )^2  + |d(\vec{k})|^2  \Big \rbrace}
\Big ]\,.
\nn
\end{align}
%%%%%%%%%%%
The first term, after expanding in small $q$, gives
\begin{align}
t_1&\propto  - e^4
 \int d^dk   \,\frac{3\,m\, q^2\,  \sin^2\! \theta }{4\, k^8 }\,,
\end{align}
which does not produce a log divergent correction to the Coulomb line. 
The second term can be written as:
%%%%%%%%%%
\begin{align}
t_2&\propto    e^4
 \int \frac{  d \omega \,d^dk}  { k^{2}\, |\vec{k}-\vec{q}|^{2} }   
\frac{ \big \lbrace i\, \omega - \frac{   k   ^2}{2\,m'}+ \dvec(\vec{k }) \cdot{\boldsymbol{\Gamma}} \big \rbrace \,
\big \lbrace    i\, \omega + \frac{k^2}{2\,m'}+ \dvec(\vec{k}) \cdot{\boldsymbol{\Gamma}}\big \rbrace   }
{   \Big \lbrace  - \left ( i\, \omega   - \frac{ k^2}{2\,m'} \right )^2 + |d (\vec{k })|^2  \Big  \rbrace   \,
\Big \lbrace - \left ( i\, \omega + \frac{k^2}{2\,m'} \right )^2  + |d(\vec{k})|^2 \Big \rbrace }
\nn
&\propto    e^4
 \int \frac{d \omega \,d^dk}  { k^{2}\, |\vec{k}-\vec{q}|^{2} }   
\frac{ -\omega^2 +\frac{k^4 \left (m'^2 -m^2 \right)} {4\, m^2\,m'^2}  }
{    \left (  \omega  ^2 +\frac{k^4 \left (m'^2 -m^2 \right)} {4\, m^2\,m'^2} 
\right )^2 +    \frac{ \omega^2 \,k^4 }{ m'^2}     } \propto   - e^4\, m'
 \int \frac{dk}  { k^{5-d }\, |\vec{k}-\vec{q}|^{2} }   \,,
\end{align}
which again does not produce a log divergent correction to the Coulomb line.

\bibliography{disorder}
%============================================================================
%.............................................................................
%======================== END DOCUMENT +++++++++++++++++++++++++++++++++++++++
\end{document}